\documentclass[amsmath,amssymb,nofootinbib,superscriptaddress]{revtex4-1}

\usepackage{aas_macros,esint,graphicx}
\usepackage{amsmath}
\usepackage{tensor} 
\usepackage{soul}
\usepackage[normalem]{ulem}
\usepackage{comment}
\usepackage{footmisc}
\usepackage[compact]{titlesec}         
\titlespacing{\subsection}{10pt}{10pt}{10pt}
\titlespacing{\section}{10pt}{10pt}{10pt}

\usepackage[allcolors=blue,colorlinks=true]{hyperref}
\numberwithin{equation}{section}

\renewcommand{\thesection}{\arabic{section}}

\makeatletter
\renewcommand{\p@subsection}{}
\renewcommand{\p@subsubsection}{}
\makeatother

\newcommand{\ed}{\mathop{}\!\mathrm{d}}

\newcommand{\nn}{\nonumber}

\usepackage{color}
\definecolor{darkgreen}{rgb}{0,0.5,0}

\begin{document}

\title{\vspace*{40pt}\huge{Quasinormal Corrections to Near-Extremal \\Black Hole Thermodynamics}\vspace*{20pt}}

\author{Daniel Kapec}
\email{danielkapec@fas.harvard.edu}
\affiliation{Center for the Fundamental Laws of Nature, Harvard University, Cambridge, MA 02138, USA}
\author{Y.T. Albert Law}
\email{ytalaw@stanford.edu}
\affiliation{Stanford Institute for Theoretical Physics, 382 Via Pueblo, Stanford, CA 94305, USA}
\author{Chiara Toldo}
\email{chiaratoldo@fas.harvard.edu}
\affiliation{Center for the Fundamental Laws of Nature, Harvard University, Cambridge, MA 02138, USA}
\affiliation{Dipartimento di Fisica, Universita' di Milano, via Celoria 6, 20133 Milano MI, Italy}
\affiliation{INFN, Sezione di Milano, Via Celoria 16, I-20133 Milano, Italy}

\begin{abstract}
\vspace*{40pt}
Recent work on the quantum mechanics of near-extremal non-supersymmetric  black holes has identified a characteristic $T^{3/2}$ scaling of the low temperature black hole partition function. This result has only been derived using the path integral in the near-horizon  region and relies on many assumptions. We discuss how to derive the $T^{3/2}$ scaling for the near-extremal rotating BTZ black hole from a calculation in the full black hole background using the Denef-Hartnoll-Sachdev (DHS) formula, which expresses the 1-loop determinant of a thermal geometry in terms of a product over the quasinormal mode spectrum. We also derive the spectral measure for fields of any spin in Euclidean BTZ and use it to provide a new proof of the DHS formula and a new, direct derivation of the BTZ heat kernel. The computations suggest a path to proving the $T^{3/2}$ scaling for the asymptotically flat 4d Kerr black hole.
\end{abstract}

\maketitle

\clearpage

{
  \hypersetup{linkcolor=black}
  \tableofcontents
}

\vspace{30pt}

\section{Introduction}

In the standard interpretation of black hole thermodynamics, the exponential of the Bekenstein Hawking entropy $e^{S_{BH}}$ represents a coarse grained entropy associated to a black hole Hilbert space. For many questions, the leading approximation to the black hole density of states, as computed  using the Euclidean black hole saddle, is sufficient. 

A notable exception occurs at low temperatures for charged or rotating non-supersymmetric black holes, which naively exhibit a large entropy $S_0$ at zero temperature. As pointed out long ago \cite{Preskill:1991tb}, for these black holes the specific heat, as computed at leading order in perturbation theory,  becomes order one at low temperatures even if the black hole itself remains macroscopic. In this case, a subleading calculation is necessary in order to determine the true low temperature phase of the black hole. As noted by these authors, one possibility is that the black hole spectrum actually has a gap, below which thermodynamics obviously does not apply. Another possibility is that the extremal ground state degeneracy is lifted at subleading order in perturbation theory, and that the states actually fill out a dense energy band above the vacuum. In this case the statistical description would have an extended range of validity at lower temperatures. Since we only have a coarse approximation to the black hole density of states through the leading saddle, either option is possible and could be realized for the black holes in our universe. 

\begin{figure}[h]
\includegraphics[scale=.15]{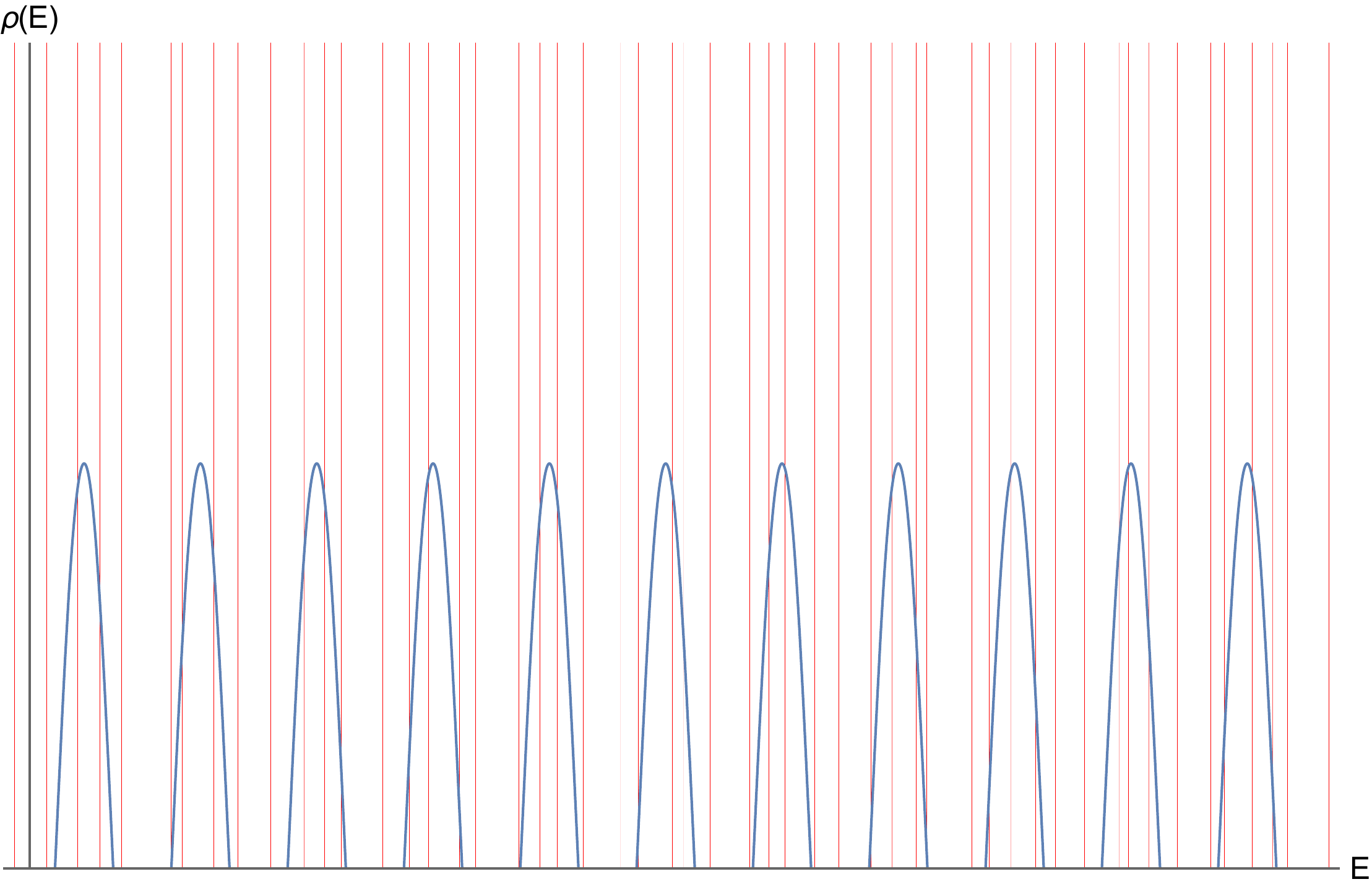}
\includegraphics[scale=.15]{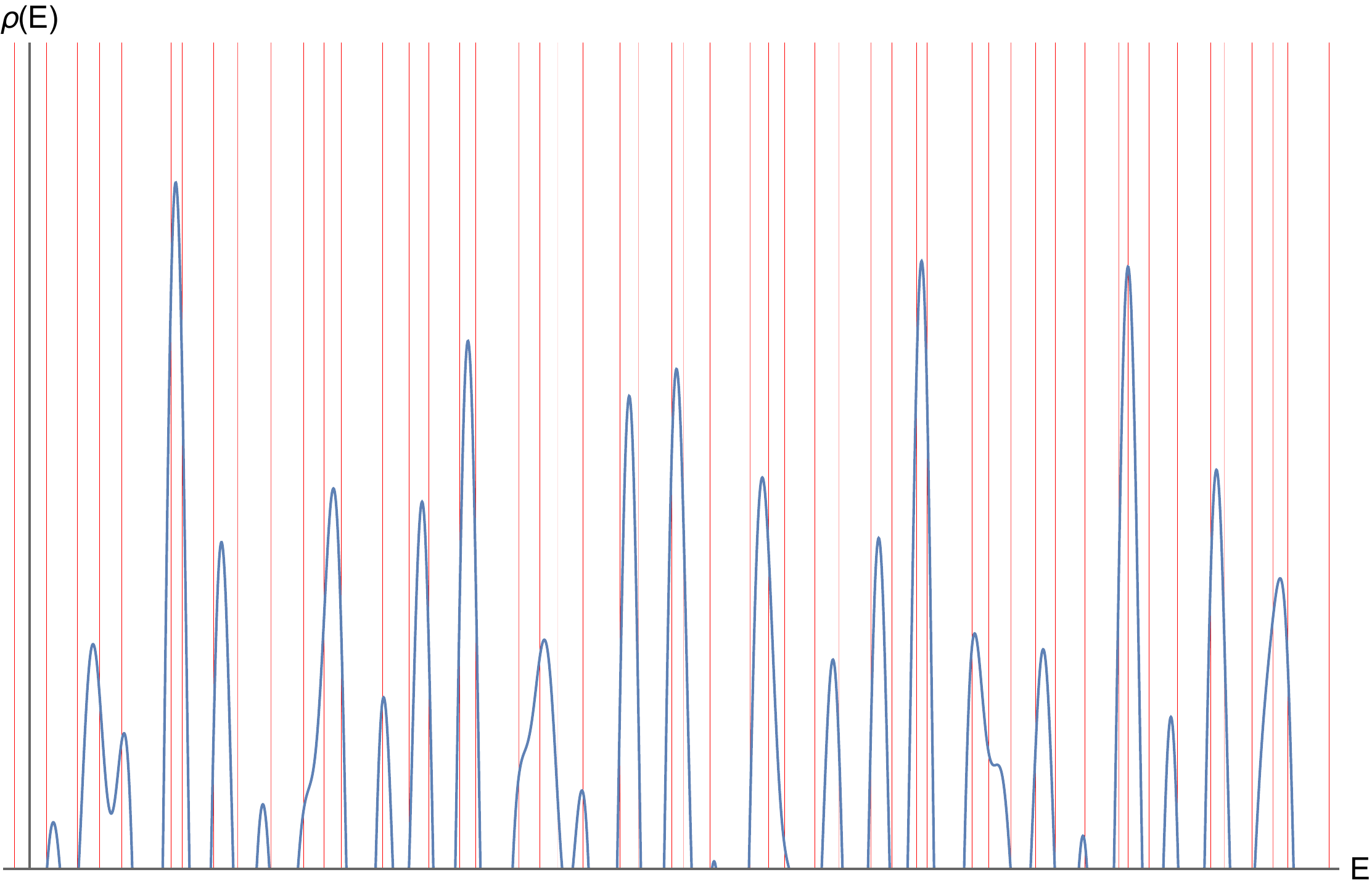}
\includegraphics[scale=.15]{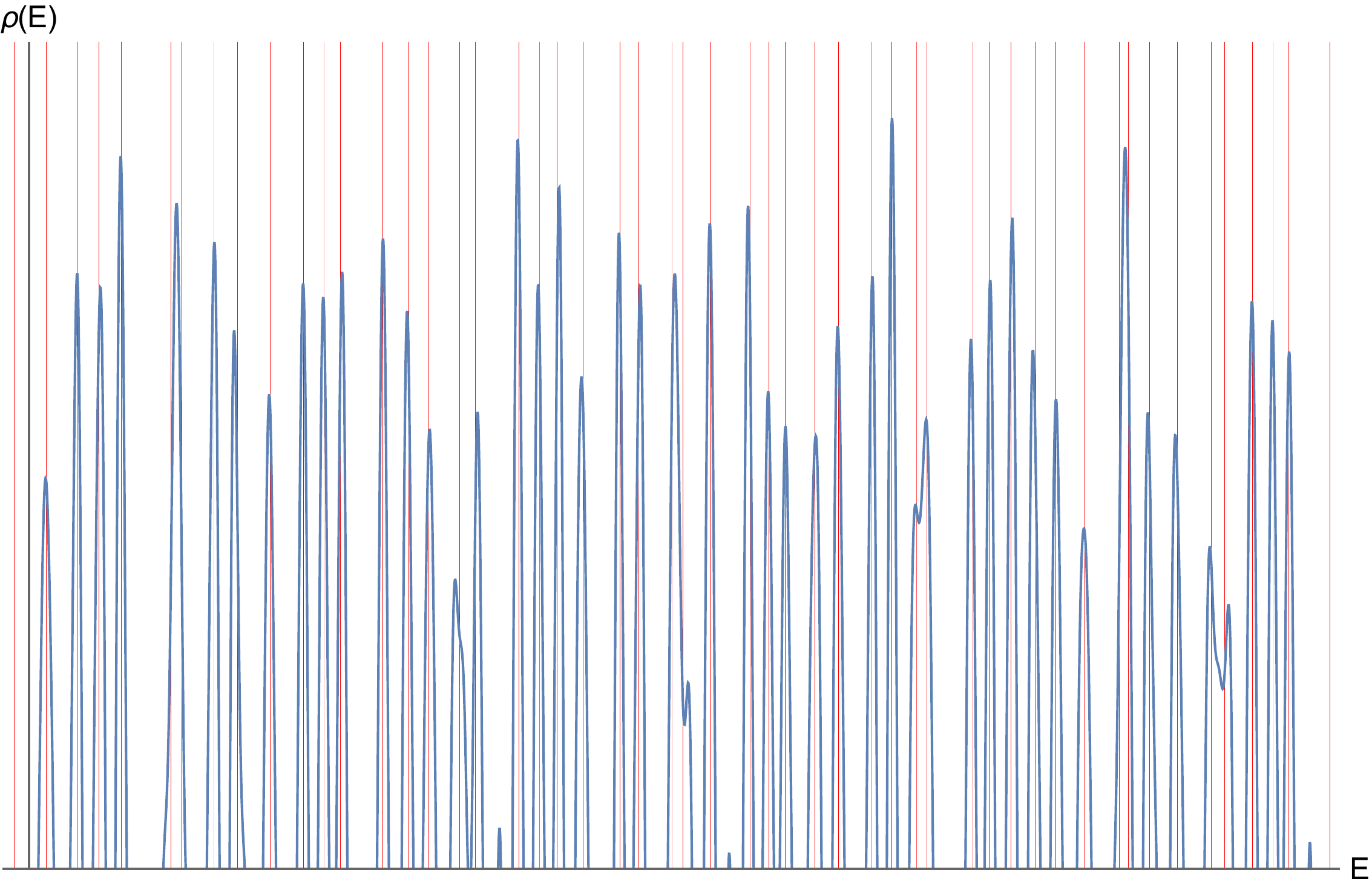}
\caption{Successive approximations (in blue) to the black hole density of states (in red) could reveal a gap in the spectrum or a dense energy band above the vacuum and no extremal ground states. }
\end{figure}

The authors of \cite{Preskill:1991tb} were not able to determine which of these two possibilities is realized in nature because they were not able to reliably calculate the 1-loop correction to the black hole partition function. According to the standard rules of Euclidean quantum gravity, this would correspond to performing the path integral over Euclidean metrics with specified boundary conditions imposed at spatial infinity
\begin{equation}\label{eq:Zfull}
Z(\beta,\mu,\Omega)\quad=    \underbrace{\int [Dg]\, e^{-I_{EH}-I_{GH}-I_{ct}}}_{\substack{\text{Asymptotically flat metrics with }\\ (\beta,\mu,\Omega) \text{ boundary conditions at } i^0}}\sim \quad  \frac{1}{\sqrt{ \det\nabla^2}}\exp{\left[-I_{\text{on-shell}}(\beta,\mu, \Omega)\right]} \; ,
\end{equation}
where $I_{EH}$ and $I_{GH}$ are the Einstein-Hilbert and Gibbons-Hawking terms and $\det \nabla^2$ represents a ratio of determinants for the graviton fluctuations as well as the  ghosts.  The exponential factor comes from the on-shell action of the black hole saddle, while the determinants arise from the integral over normalizable fluctuations about the saddle-point. Because the exponential term is so large, the tree level calculation typically dominates the thermodynamics. In other words, we view the determinant as a small correction which in most circumstances does not change the qualitative behavior of the thermodynamic system. This is certainly true for a gas of free gravitons in flat space, whose thermodynamic behavior is relatively standard (setting aside known subtleties regarding the conformal factor and negative modes above the Davies point \cite{Davies:1977bgr,Monteiro:2008wr}) but recent work \cite{Iliesiu:2020qvm,Heydeman:2020hhw,Boruch:2022tno,Iliesiu:2022kny,Iliesiu:2022onk,Banerjee:2023quv,Banerjee:2023gll,Kapec:2023ruw,Rakic:2023vhv} has indicated that a gas of gravitons at low temperatures in a black hole background can dramatically change the thermodynamic behavior of the system. It is this correction which determines the answer to the questions raised in \cite{Preskill:1991tb}.

Since calculating and regulating the determinants in \eqref{eq:Zfull} requires expressions for the eigenvalues and degeneracies of the kinetic operator in the curved black hole geometry, the 1-loop correction typically cannot be calculated explicitly. The full calculation is certainly out of reach for the Kerr black hole, and that is the reason that the authors of \cite{Preskill:1991tb} could not offer a definitive answer. The recent progress on this problem proceeds from the observation (more accurately, the assumption) that there is an alternate way to compute the correction to the low temperature thermodynamics that involves only fluctuations in the near-horizon throat region of the black hole.  This ability to replace the full black brane geometry with its near-extremal throat (for appropriate low energy observables) lies at the heart of the AdS$_{d+1}$/CFT$_d$ correspondence, but its application to black holes with $d=1$ is notoriously subtle. 

 According to the AdS/CFT dictionary, instead of performing the full asymptotically flat path integral \eqref{eq:Zfull} and isolating the contribution from the black hole, one can equivalently calculate a quantity
\begin{equation}\label{eq:Zthroat}
Z_{\text{throat}}(\beta,\mu,\Omega)\quad=    \underbrace{\int [Dg]\, e^{-I_{EH}-I_{GH}-I_{ct}}}_{\substack{\text{Asymptotically AdS}_{d+1}\text{ metrics with }\\ (\beta,\mu,\Omega) \text{ boundary conditions at } \partial \text{AdS}}} \; 
\end{equation}
for sufficiently low temperatures with macroscopically long throats. This is thought to represent a trace over the black brane Hilbert space, which is assumed to decouple from the far region for sufficiently low temperatures. The dictionary relates the left hand side to the thermal partition function of an ordinary quantum system, usually a strongly coupled low energy limit of strings and branes. The right-hand-side is thought to contain valuable information about the thermodynamics of this strongly coupled quantum system, although often only the leading saddle point approximation is considered.

In the present context, one would like to use the 1-loop approximation to \eqref{eq:Zthroat}, which is substantially simpler than \eqref{eq:Zfull} due to enhanced symmetries in the throat, to answer the question posed in \cite{Preskill:1991tb}. 
 Unfortunately it is precisely the case when $d=1$ that \eqref{eq:Zthroat} is most subtle.  A strict zero temperature version of \eqref{eq:Zthroat} was used by Sen and collaborators to calculate logarithmic corrections to extremal black hole entropy \cite{Sen:2007qy,Sen:2008vm,Sen:2009vz,Sen:2012dw,Sen:2012cj,Sen:2012kpz,Sen:2014aja}. Zero temperature partition functions count ground states, so these authors really calculated a particular correction to the quantity
\begin{equation}\label{eq:Zlog}
    Z_{\text{throat}}(\beta=\infty,Q,J)=\text{Tr}_{\text{groundstates}}\sim S_0^{c_{\text{log}}}e^{S_0} \; .
\end{equation}
Here, $c_{\text{log}}$ is a universal constant fixed by the low energy field content of the model, $S_0$ is the extremal Bekenstein-Hawking entropy arising from the tree level calculation, and we have suppressed counterterm dependent quantities that relate to renormalization of the ground state energy. It is important to note that the calculation is performed at fixed charge, rather than fixed chemical potential, because the non-normalizable component of the AdS$_2$  gauge field differs from the non-normalizable component in the four-dimensional asymptotically flat far region. 

In supersymmetric examples the calculation of \eqref{eq:Zlog} is robust, but there are known difficulties in using \eqref{eq:Zthroat} to account for finite energy excitations \cite{Maldacena:1998uz,Amsel:2009ev} when $d=1$.  
This is because the extremal throat of black holes does not quite decouple from the far region in the low energy limit. In Lorentzian language, the throat has low dimensionality, so the gravitational field of perturbative excitations does not have enough directions to spread out in. It  thus grows towards the mouth of the throat where it interacts with the far region, so that generic finite energy excitations  violate the strict AdS$_2$ asymptotics.

This problem manifests itself in the Euclidean calculation via the presence of infinitely many normalizable zero modes, corresponding to reparametrizations of boundary time, in the extremal throat. This leads to an infrared divergence in the path integral \eqref{eq:Zthroat} at zero temperature
\begin{equation}\label{eq:Zinf}
    Z_{\text{throat}}\propto \int \displaylimits_{\text{Diff}(S^1)/\text{SL}(2,\mathbb{R})}[Dh] \;  = \infty \; .
\end{equation}
Sen extracted the dependence of the measure on $S_0$ in order to calculate \eqref{eq:Zlog}, but the path integral itself is ill-defined.

There was recent progress on this problem in the case of spherically symmetric
\cite{Iliesiu:2020qvm,Heydeman:2020hhw,Boruch:2022tno,Iliesiu:2022kny,Iliesiu:2022onk,Banerjee:2023quv,Banerjee:2023gll} and rotating black holes \cite{Kapec:2023ruw,Rakic:2023vhv}. The basic idea of these analyses is to regulate the infrared divergence \eqref{eq:Zinf} by maintaining the subleading correction to the throat metric in the near-extremal limit. In other words, one performs the usual near-horizon near-extremal scaling limit in Boyer-Lindquist $(\hat{t},\hat{r},\theta,\hat{\phi})$ coordinates
\begin{align}\label{eq:Scalinglimit}
	\hat{t}=\frac{1}{2\pi T}t,\qquad
	\hat{r}=r_+(T)+4\pi r_0^2 T  (r-1),\qquad
        \hat{\phi}=\phi +\frac{t}{4\pi r_0 T}-t	\; ,
\end{align}
whose strict $ T\to 0$ limit leads to the NHEK metric
\begin{equation}\label{eq:NHEK}
	ds^2_{\text{NHEK}}=J(1+\cos ^2\theta)\left(-(r^2-1)\ed t^2+\frac{\ed r^2}{r^2-1}+\ed\theta^2\right)+
 J\frac{4\sin^2\theta}{1+\cos^2\theta}\left(\ed\phi+(r-1)\ed t\right)^2 \; .
\end{equation}
However, contrary to what is usually done, one retains the subleading term in the small $T$ expansion
\begin{equation}\label{eq:IntroNotNHEK}
    g_{\text{not-NHEK}}=g_{\text{NHEK}}+  T\delta g \; .
\end{equation}
The resulting ``not-NHEK'' metric is not diffeomorphic to the NHEK metric and is not an exact solution to the Einstein equation. Rather, it can be viewed as an irrelevant deformation of the NHEK metric whose nonlinear completion is the finite 
(but low) temperature Kerr black hole.\footnote{Note that there are infinitely many irrelevant deformations one could perform in the NHEK throat, roughly corresponding to solutions with extra dynamics away from the horizon, each of which would presumably also regulate the IR divergence \eqref{eq:Zinf}.} Performing the path integral about this new background metric yields a regulated version of \eqref{eq:Zinf}
\begin{equation}\label{eq:Zreg}
Z_{\text{reg}}(\beta,Q,J)\quad=    \underbrace{\int [Dg]\, e^{-I_{EH}-I_{GH}-I_{ct}}}_{\substack{\text{Asymptotically ``not-NHEK'' metrics with }\\ (\beta,Q,J) \text{ boundary conditions at } \partial \text{(not-NHEK)}}}\; 
\end{equation} 
 The path integral becomes finite because the zero modes of the extremal throat are large diffeomorphisms of NHEK 
\begin{equation}\label{eq:zeroModes}
    h^{(n)}=\mathcal{L}_{\xi^{(n)}}g_{\text{NHEK}} \; , \qquad \qquad \xi^{(n)} \quad \text{non-normalizable}
\end{equation}
but are not Lie derivatives of the ``not-NHEK'' metric
\begin{equation}\label{eq:zmnotNHEK}
h^{(n)}\neq\mathcal{L}_{\zeta}g_{\text{not-NHEK}} \; .
\end{equation}
They therefore pick up nonzero eigenvalues at first order in perturbation theory, and one finds that their contribution to the zeta regularized determinant 
\begin{equation}\label{eq:ZetaReg}
    \prod_{n\geq 2}\frac{1}{nT}=\frac{1}{\sqrt{2\pi}}T^{3/2}
\end{equation}
exhibits the characteristic $T^{3/2}$ scaling. Since the regularized partition function becomes small at low temperatures 
\begin{equation}\label{eq:smallT}
    Z_{\text{reg}}\sim T^{3/2}e^{S_0}
\end{equation}
one concludes that the ground states are lifted, provided that it is correct to interpret the regulated path integral   \eqref{eq:Zreg} as computing the black hole partition function at small finite temperature.

Unfortunately, while plausible, this prescription has not really been derived. The question of whether or not to integrate over diffeomorphisms with non-compact support depends delicately on boundary conditions and gauge fixing, and one also has to consider the ambiguities in connecting the far region to the throat. The modes which are (non)normalizable in the throat are not guaranteed to complete to (non)normalizable modes in the full asymptotically flat geometry (e.g. the proceeding discussion regarding the different ensembles in \eqref{eq:Zfull} and \eqref{eq:Zlog}). Relatedly, the zero-mode metric perturbations \eqref{eq:zeroModes} arising from diffeomorphisms with non-compact support in AdS$_2$/NHEK might correspond to pure diffeomorphisms with compact support in the full Kerr geometry, or they might complete to physical non-zero modes in Kerr which simply reduce  to diffeomorphisms in the throat region. The latter should be the case if \eqref{eq:smallT} is to be a reliable approximation to \eqref{eq:Zfull} at low temperatures.

Given these apparent subtleties, one would like to reproduce the behavior \eqref{eq:smallT} using the full asymptotically flat geometry and verify that the contribution of the zero modes to the throat calculation is physical. At first glance this seems out of reach, since we cannot even perform the full not-NHEK path integral, only the piece responsible for the $T^{3/2}$ behavior. The calculation in near-extremal Kerr appears hopelessly more complicated.

However, there is an interesting formula due to Denef, Hartnoll and Sachdev (DHS) which expresses the Euclidean partition function in terms of the Lorentzian quasinormal  modes (QNMs) \cite{Denef:2009kn,Castro:2017mfj}
\begin{align}
	Z_{\rm PI} = \prod_{k,l\in\mathbb{Z}} \prod_{z_l}\left( \omega_{|k|,l}+ i z_l\right)^{-1/2} \;.
\end{align}
In this formula the $\omega_{k,l}=2\pi k T -il\Omega$ are the Matsubara frequencies for the ensemble with temperature $T$ and angular velocity $\Omega$, fixed by regularity and periodicity on the Euclidean section of the black hole geometry. The $z_{l}$ are the QNMs for the field whose determinant we are computing.

It is not obvious that this formula simplifies the problem since we cannot analytically compute the full quasinormal spectrum for black holes in more than three dimensions. However, there is a particular branch of  QNMs whose frequencies can be computed analytically and which are closely related to the existence of the throat in the near-extremal Kerr geometry. 
These ``lightly-damped'' modes \cite{Detweiler:1980gk,Sasaki:1989ca,Glampedakis:2001fg,Ferrari:1984zz,Cardoso:2004hh,Hod:2008zz,Hod:2012bw,Yang:2012he,Yang:2012pj,Yang:2013uba,Cook:2014cta,Dias:2015wqa} have real parts that accumulate at the superradiant bound and small imaginary parts spaced evenly in units of the Hawking temperature 
\begin{equation}\label{eq:Spec}
    \omega=m\Omega_H -2\pi i T_H \left(n+\frac12\right) \; .
\end{equation}
The imaginary parts are small precisely because waves with $\omega=m\Omega_H$ enter and spend a long time in the throat region, so that they have long lifetimes. Indeed, \eqref{eq:Spec} is really the Lorentzian spectral signature of the existence and enhanced conformal symmetry of the NHEK throat \cite{Hadar:2022xag,Kapec:2022dvc}, and one might expect that these modes have something to do with the low temperature corrections \eqref{eq:smallT} since they are most sensitive to the throat geometry.

The idea then would be to identify and separate out the contribution of the throat region to the black hole determinant using the DHS formula, discarding the rest of the terms in the product whose QNMs don't really have anything to do with extremality
\begin{equation}\label{eq:DHS32}
    Z(\beta,\Omega)= \left[\prod_{{\substack{\text{throat } \\ \text{contribution}}}}\right] \left[\prod_{\text{all other QNM}    }\right] \sim T^{3/2}\left[\prod_{\text{all other QNM}    }\right] \; ?
\end{equation}
The second incalculable term will correspond to a non-universal contribution which does not have a singular limit as $T\to 0$ since it is not really sensitive to the geometry near the throat. A related idea was recently explored in \cite{Mukherjee:2024nhx}, focusing instead on reproducing the logarithmic corrections to the precisely extremal black hole entropy rather than the low temperature corrections considered herein, and it would be interesting to combine the two calculations.

As a warm-up for the Kerr problem, in this paper we perform the analogous calculation for the near-extremal BTZ black hole. In this case, we know the QNM spectrum exactly \cite{Datta:2011za}, the 1-loop Euclidean determinant independently \cite{Maloney:2007ud,Giombi:2008vd}, and the relevant limit of the corresponding CFT$_2$ character which reproduces the $T^{3/2}$ behavior \cite{Ghosh:2019rcj}. 
Indeed, the DHS formula has already been applied in this case \cite{Castro:2017mfj, Keeler:2018lza,Keeler:2019wsx} although the low temperature limit was not investigated. Taking the low temperature limit of the determinant in Lorentzian variables allows us to isolate the specific graviton  QNMs responsible for the $T^{3/2}$ scaling in the full black hole geometry. It turns out to be crucial that certain QNMs for spinning fields do not continue to regular Euclidean solutions with low Matsubara frequencies \cite{Datta:2011za,Castro:2017mfj,Grewal:2022hlo}.  As we show, their exclusion from the DHS product formula is intimately related to the low-temperature scaling, playing the same role that the exact $SL(2,\mathbb{R})$ symmetry plays in the throat calculation. 

We are optimistic that finding the analogous statements for the Kerr determinant and performing similar manipulations will provide an unambiguous  proof of the $T^{3/2}$ scaling for near-extremal Kerr. Of course, the BTZ  QNM spectrum is dramatically simpler than that of Kerr (although it does closely resemble the particular branch \eqref{eq:Spec}), so the exact combinations of terms producing the $T^{3/2}$ may differ significantly. Rotating BTZ also has the added simplification that it does not exhibit superradiance, which we expect to complicate the analysis in Kerr as described in \cite{Kapec:2023ruw}.

A particularly interesting byproduct of our analysis is an explicit formula (derived in section \ref{sec:BTZ}) for the spectral measure for fields of arbitrary spin in Euclidean BTZ. We use this expression to provide a new proof of the DHS formula and new, independent derivations of the BTZ heat kernel for fields of all integer spin. The fact that we are able to reproduce the known results for determinants using only a continuous integral over eigenvalues seems to support the conclusion that there is no discrete spectrum for the relevant kinetic operators in Euclidean BTZ, in accord with the conclusions in \cite{Acosta:2021oqt}. This would mean that the product form of the DHS formula really is the only way to make contact with the discrete product \eqref{eq:ZetaReg} encountered in the throat calculation using the full black hole geometry. We must replace the discrete product of eigenvalues in the throat by a discrete product of  QNMs in Kerr.

The outline of this paper is as follows. Section \ref{sec:throat} derives the $T^{3/2}$ scaling using the deformed throat geometry of near-extremal BTZ and a discrete product over nearly-zero modes, mirroring the analysis of \cite{Kapec:2023ruw,Rakic:2023vhv}. In section \ref{sec:extremalZ} we take the extremal limit of the BTZ partition function as in \cite{Ghosh:2019rcj}, but in a way that makes manifest the connection to the calculation in section \ref{sec:throat}. Section \ref{sec:DHS} provides a compact derivation of the BTZ determinant for all spins using the DHS formula and the techniques of \cite{Anninos:2020hfj,Law:2022zdq,Grewal:2022hlo}. In section \ref{sec:T32} we discuss which specific graviton QNMs are responsible for the $T^{3/2}$ scaling in the low temperature limit. In section \ref{sec:BTZ}  we derive the spectral measure for spinning fields in Euclidean BTZ and use it to explicitly prove the DHS formula for the determinant. This also allows us to provide an independent derivation of the previously known BTZ heat kernel. Section \ref{sec:Conclusion} concludes with comments on the possible generalization to 4d Kerr and Reissner-Nordstrom black holes. Appendix \ref{app:HS} discusses low-temperature $\log T$ corrections arising from massless higher spin fields in BTZ.

After our results were finalized and being written up, we became aware that Kolanowski, Rakic, Rangamani, and Turiaci are also considering the origin of the $T^{3/2}$ scaling for the BTZ black hole.

\section{The deformed BTZ throat and $T^{3/2}$ scaling }\label{sec:throat}
In this section we rederive the results of \cite{Kapec:2023ruw,Rakic:2023vhv} for the simpler case of the near-extremal rotating BTZ black hole.
The Lorentzian rotating BTZ geometry 
\begin{align}\label{eq:BTZBLmetric}
	 ds^2 = -\frac{(r^2 - r_+^2)(r^2 - r_-^2)}{ \ell^2r^2} dt^2 +  \frac{\ell^2r^2}{(r^2 - r_+^2)(r^2 - r_-^2)}dr^2 + r^2 \left( d \varphi - \frac{r_+ r_-}{\ell r^2 } dt \right)^2 
\end{align}
is a solution to 3d gravity with a negative cosmological constant
\begin{equation}
 I_{3d}=   \frac{1}{16\pi G}\int_M d^3x \sqrt{g} \left(R+\frac{2}{\ell^2}\right) -\frac{1}{8\pi G} \int_{\partial M}d^2x \sqrt{\gamma}K  \; + I_{ct}\; . 
\end{equation}
The inner and outer radii are given by
\begin{equation}
    r_{\pm}^2= \frac{M\ell^2}{2}\left(1\pm \sqrt{1-\frac{J^2}{M^2\ell^2}}\right) \; . 
\end{equation}
Equivalently, the black hole mass and spin are given by
\begin{align}
	M = \frac{r_+^2 + r_-^2}{\ell^2}\; ,\qquad \qquad	J = \frac{2r_+ r_-}{\ell}\;. 
\end{align}
The entropy and temperature are
\begin{equation}
    S_{\text{BH}}=\frac{2\pi r_+}{4G} \; ,  \qquad T = \frac{r_+^2-r_-^2}{2 \pi r_+ \ell^2} \; . 
\end{equation}
For completeness, we also introduce the left and right moving temperatures
\begin{equation}\label{leftright}
     T_{L}= \frac{r_+ - r_-}{2\pi} \; , \qquad T_{R}= \frac{r_+ + r_-}{2\pi} \;  
\end{equation}
that will be useful in later sections.
Expanding the entropy about low temperatures one finds (setting $G=1$)
\begin{equation}
    r_+(T,J)=\sqrt{\frac{J\ell}{2}} + \frac{1}{2}\pi \ell^2 T + O(T^2) \; , \qquad \qquad  S_{BH}(T,J)=\frac{\pi}{2}\sqrt{\frac{J\ell}{2}} + \frac{\pi^2\ell^2}{4} T + O(T^2) \; .
\end{equation}
Similarly, the mass has a small-$T$ expansion
\begin{eqnarray}
	M(T,J) & = & \frac{J}{\ell} +  \ell^2 \pi^2 T^2 + \frac{ \sqrt2 \ell^4 \pi^3 T^3}{ \sqrt{J \ell} } + \frac{3 \ell^5 \pi^4 T^4}{2J} + O(T^5) \; . 
\end{eqnarray}
These formulas differ in an important way from the analogous formulas in  Kerr, where 
\begin{equation}
    S_{\text{Kerr}}(T,J)=S_0+8\pi^2 J^{3/2}T + O(T^2) \; , \qquad \qquad M_{\text{Kerr}}(T,J)=M_0 + 4\pi^2J^{3/2}T^2+ \cdots \; . 
\end{equation}
In particular, the ``gap scale'' no longer depends on the angular momentum and is simply fixed by the AdS$_3$ radius.

With this in mind, we perform the following diffeomorphism
\begin{equation} \label{eq:BTZ_coord_eu}
   r = r_+ + \frac{T}{2} \left( \ell^2 \pi  (z-1) \right)\; ,  \qquad t =  \frac{\tau}{2\pi  T} \; , \qquad \varphi = \ell  \phi + \tau \left(\frac{1}{2 \ell \pi T} - \frac{\ell}{2r_0} \right)\;, \qquad T\to 0 \; , 
\end{equation}
were $r_0^2=\frac{\ell}{2}J$, obtaining
\begin{equation} \label{throatBTZ_lor}
   ds^2_{\text{throat}} = \frac{\ell^2}{4} \left(1 - z^2 \right) d\tau^2 +\frac{\ell^2 dz^2 }{4(z^2 - 1)} + \ell^2 r_0^2\left(d\phi - \frac{1}{2r_0}\left(1-z \right) d\tau \right)^2 \; . 
\end{equation}
The metric \eqref{throatBTZ_lor} is the near-horizon geometry of BTZ. It can be viewed as a discrete quotient of AdS$_3$, known as the ``self-dual orbifold" \cite{Coussaert:1994tu,Balasubramanian:2003kq}. Importantly, the discrete group relating extremal BTZ to a quotient of AdS$_3$ is inequivalent \cite{Balasubramanian:2003kq} to the quotient group for the self-dual orbifold: extremal BTZ is not globally diffeomorphic to its near-horizon geometry. 
Wick rotating $\tau \rightarrow -i \bar{t}$ and changing to the radial coordinate  $z = \cosh \eta$
we obtain the Euclidean throat geometry
\begin{equation}\label{BTZ_coord_ext}
   ds^2_{\text{throat}} = \ell^2 \sinh^2\left(\frac{\eta}{2}\right) d\bar{t}^2 +\frac{\ell^2}{4} d\eta^2+i \ell^2 r_0(1-\cosh \eta) d\bar{t} d\phi + \ell^2 r_0^2 d\phi^2 \; .
\end{equation}
 Linearizing the Lagrangian (plus a gauge fixing term) around this solution, the path integral is  \cite{Giombi:2008vd}
\begin{equation}
  Z \sim \exp( -I_{3d}[\overline{g}]) \int [Dh] \exp\left[ -\int d^3x \sqrt{\bar{g}} h D[\bar{g}] h \right]\; ,  \quad  h_{\mu \nu }D_{BTZ}^{\alpha\beta, \mu \nu} h_{\alpha \beta} =  -\frac{1}{64 \pi} h_{\mu \nu} (g^{\mu \alpha} g^{\nu \beta} \nabla^2 + 2R^{\mu \alpha \nu \beta} ) h_{\alpha \beta} \; . 
\end{equation}
The metric \eqref{BTZ_coord_ext} admits normalizable zero modes (transverse traceless perturbations) of the form 
\begin{equation}\label{eq:throatZeroMode}
    h_{\mu \nu}^{(n)} dx^\mu dx^\nu= \frac{\sqrt{2\ell}}{ 4 \pi} \sqrt{\frac{ |n|(n^2-1)}{r_0 }} \frac{e^{in \bar{t}}(\sinh \eta)^{|n|-2}}{(1+\cosh \eta)^{|n|}}    (d\eta^2+2i\frac{n}{|n|} \sinh \eta d\eta d\bar{t}-\sinh^2\eta d\bar{t}^2) \; , \quad |n|>1 \; . 
 \end{equation}
In order to regulate the infrared divergence we keep the subleading correction in the decoupling limit \eqref{eq:BTZ_coord_eu}
\begin{equation} \label{BTZ_coord_eu}
   g = g_{\text{throat}} + T  \delta g
\end{equation}
with the irrelevant deformation given by
\begin{eqnarray} \label{eq:notNHEK}
\delta g_{\mu \nu}dx^\mu dx^\nu & = & \frac{\ell^4 \pi }{2r_0} \sinh^4\left(\frac{\eta}{2}\right)d\bar{t}^2 +\frac{ \ell^4 \pi  }{8r_0} (2 + \cosh \eta) \tanh^2 \left(\frac{\eta}{2} \right)d\eta^2  + \ell^4\pi r_0  \cosh \eta d\phi^2  \nonumber \\
& - &i\frac{\ell^4\pi}{2}   ( \cosh \eta-3)  \sinh^2\left( \frac{\eta}{2} \right) d\bar{t} d\phi  \; .
\end{eqnarray}
Applying first order perturbation theory, the eigenvalues of the zero modes \eqref{eq:throatZeroMode} become
\begin{equation}\label{eq:correction}
    \delta \Lambda_n = \int d^3 x \sqrt{\bar{g}} h^{(n)}_{\alpha \beta} \delta D^{\alpha \beta , \mu \nu} h^{(n)}_{\mu \nu} \; , 
\end{equation}
and the correction to the determinant is
\begin{equation}\label{eq:ThroatSum}
    \log Z_{\text{throat}} = - \frac12 \sum_n \log ( \delta \Lambda_n) \;. 
\end{equation}
The real and imaginary parts of \eqref{eq:throatZeroMode}
have the same eigenvalue correction
\begin{equation}
    \int d^3x \sqrt{\bar{g}} h_{\alpha \beta }^{(n)}\delta D^{\alpha\beta, \mu \nu} h^{(n
)}_{\mu \nu}= \frac{n T}{32r_0} 
\end{equation}
so the  contribution to the BTZ throat partition function is 
\begin{equation} \label{eq:ZeroModeProd1}
\delta \log Z_{\text{throat}} = 2 \left(-\frac12\right) \sum_{n \geq 2} \delta \Lambda_n = \log \left( \prod_{n \geq 2} \frac{32r_0}{n T}  \right)\; .  
\end{equation}
 Using zeta function regularization
\begin{equation}\label{eq:ZeroModeProd}
    \prod_{n>1}^{\infty} \frac{c}{n} = \frac{1}{\sqrt{2 \pi}} \frac{1}{c^{3/2}} 
\end{equation}
we obtain 
\begin{equation}\label{eq:throatT32}
    \delta \log Z_{\text{throat}} = \log \left( \frac{ 1}{ 256\sqrt{\pi}}\left(\frac{T}{r_0}\right)^{3/2} \right) \sim \frac32 \log T 
\end{equation}
which exhibits the familiar $\frac32 \log T$ behaviour, as in \cite{Kapec:2023ruw,Rakic:2023vhv}.

\section{Near-extremal limit of the BTZ path integral }\label{sec:extremalZ}

Because the Euclidean BTZ geometry is a modular transform of thermal AdS$_3$, the BTZ 1-loop path integral can be obtained from the 1-loop path integral about the thermal AdS$_3$ saddle. The thermal AdS$_3$ path integral can in turn be interpreted as a (bulk) trace, whose form is constrained by the Virasoro $\times$ Virasoro asymptotic symmetry group of AdS$_3$ quantum gravity \cite{Brown:1986nw}. In particular, for a scalar field of mass $m^2$ in thermal AdS with boundary torus modular parameter $\tau$ the one-loop determinant takes the form \cite{Maloney:2007ud,Giombi:2008vd}
\begin{equation}\label{eq:scalarDet}
\frac{1}{\det (-\nabla_{s=0}^2+m^2)^{1/2}} =   \prod_{n,n'=0}^\infty \frac{1}{1-q^{n+h}\bar{q}^{n'+h}} \; , \qquad q=e^{2\pi i \tau} \; ,\qquad  h = \frac12 \left(1+\sqrt{1+m^2}\right) \; . 
\end{equation}
This can be rewritten as a trace over $SL(2,\mathbb{R})\times SL(2,\mathbb{R})$ descendant states of the form $L^{n}_{-1}\bar{L}_{-1}^{n'}|h,\bar{h}\rangle$. The analogous graviton determinant combined with the tree level calculation brings along the full family of Virasoro descendants and takes the form (suppressing scheme-dependent factors that renormalize local counterterms)
 \begin{equation}\label{eq:TAdS3Z}
     Z^{\text{graviton}}_{\text{TAdS}_3}(\tau, \bar{\tau})=\chi_1(\tau)\chi_1(\bar{\tau})\; , \qquad \qquad  \tau=\frac{i}{2\pi}\beta_L \; ,\qquad \bar{\tau}=-\frac{i}{2\pi}\beta_R \; . 
 \end{equation}
Here $\chi_1(\tau)$ is the identity character of the Virasoro algebra 
\begin{equation}
    \chi_1(\tau)=\frac{(1-q)q^{\frac{1-c}{24}}}{\eta(\tau)}\; , \qquad c=\frac{3\ell}{2G} \;  
\end{equation} 
with the Brown-Henneaux central charge. This gravity partition function without matter is one-loop exact \cite{Maloney:2007ud}. The formula \eqref{eq:TAdS3Z} is fixed by representation theory  \cite{Maloney:2007ud}  but has also been calculated explicitly \cite{Giombi:2008vd} by combining the heat kernel on hyperbolic three-space $\mathbb{H}_3$ with  the method of images (thermal AdS$_3$ and BTZ are both $\mathbb{Z}$ quotients of $\mathbb{H}_3$).

The BTZ partition function is the modular transform of \eqref{eq:TAdS3Z}
\begin{equation}
    Z_{BTZ}(\beta_L,\beta_R)=\chi_1(-1/\tau)\chi_1(-1/\bar{\tau})= \chi_1\left(\frac{2\pi i}{\beta_L}\right)\chi_1\left(-\frac{2\pi i}{\beta_R}\right)\; .
\end{equation}
The left and right temperatures are related to the temperature and angular velocity of the black hole according to 
\begin{equation}\label{eq:TlTr}
    \frac{2}{T} = \frac{1}{T_L}+\frac{1}{T_R}\; , \qquad \Omega_{H} =  \frac{r_-}{r_+}=\frac{T_R-T_L}{T_R+T_L} \; .
\end{equation}
In these formulas $T_L$ and $T_R$ are complex conjugates so that $T$ is real and $\Omega_{H}$ is purely imaginary. The near-extremal limit  corresponds to taking\footnote{ As described in \cite{Ghosh:2019rcj}, this limit requires that we continue $\Omega_H$ back to real values so that the Euclidean geometry is complex while the Lorentzian geometry is real. }
\begin{equation}\label{beta_BTZ}
    \beta_R\to 0 \; , \qquad \beta_L\to \infty \; , 
\end{equation}
or equivalently $T_L\to 0$ with $T_R\to \infty$ so that $T\sim 2T_L\ll 1$ and $\Omega\to 1$. Taking this limit \cite{Ghosh:2019rcj} using 
\begin{equation}
    \eta\left(-\frac{1}{\tau}\right)=\sqrt{-i\tau}\, \eta(\tau)\; 
\end{equation}
and the asymptotic $\eta(\tau)\to e^{\frac{i\pi}{12}\tau} $ as $ \tau \to i\infty$ one finds the leading behavior
\begin{align}\label{eq:leftAsymp}
    \chi_1\left(\frac{2\pi i }{\beta_L}\right)&\sim 2\pi \left(\frac{2\pi}{\beta_L}\right)^{3/2}\exp{\left[\frac{\beta_L}{24}+\frac{c-1}{24}\frac{(2\pi)^2}{\beta_L}\right]} \;  . 
\end{align}
In this paper we are primarily concerned with reproducing the factor $T^{3/2}$ that arises from the left moving character in \eqref{eq:leftAsymp} and will ignore exponential terms and terms that depend only on $T_R$.

When comparing to the form of the throat calculation \eqref{eq:ZeroModeProd1} a different expression for the BTZ determinant is more useful. The product representation of the identity character leads to a representation for the one loop BTZ path integral of the form
\begin{equation}\label{eq:AdSgravPI}
    Z_{BTZ}(\tau, \bar{\tau})=|q_Lq_R|^{-c/24}\prod_{n=2}^\infty \frac{1}{1-q_L^{n}}\prod_{n=2}^\infty \frac{1}{1-q_R^{n}} \;, \qquad q_L=e^{-(2\pi)^2T_L} \; , \qquad q_R=e^{-(2\pi)^2T_R} \; .  
\end{equation}
Since we know that it is the left character which accounts for the $T^{3/2}$ scaling we will ignore the right character as well as the factor $|q_Lq_R|^{-c/24}$ which arises from the tree level calculation. Then in the extremal limit
\begin{equation}\label{eq:BTZprod}
    Z_{\text{BTZ}} \sim \prod_{n=2}^\infty \frac{1}{1-\exp[-\frac{1}{2}(2\pi)^2nT]}\sim \prod_{n=2}\frac{2(2\pi)^{-2}}{nT} \; , 
\end{equation}
which is precisely the form of the determinant \eqref{eq:ZetaReg} encountered in all deformed-throat calculations. As in the throat calculation, the important fact about the infinite product is that it starts with $n=2$. In the throat calculations this exclusion results from the exact $SL(2,\mathbb{R})$ symmetry of the vacuum. Since the full BTZ geometry does not have conformal symmetry, we would like to understand how to account for the restricted range in the product using gravitational variables. As we will see, this is most clear in the DHS representation of the BTZ determinant.

\section{Quasinormal modes and the DHS formula }\label{sec:DHS}
In this section we re-derive the 1-loop correction to the BTZ path integral using the DHS formula for fields of various spin using the techniques introduced in \cite{Anninos:2020hfj,Law:2022zdq,Grewal:2022hlo}. For earlier treatments see \cite{Castro:2017mfj, Keeler:2018lza,Keeler:2019wsx}. We set $\ell_{\text{AdS}}=1$.

\subsection{Real scalar }
In this subsection we consider the one-loop determinant for a scalar field of mass $m^2=\Delta \left(\Delta -2 \right)$ on the Euclidean BTZ black hole background
\begin{align}
    Z_{\rm PI} = \frac{1}{\det (-\nabla_{s=0}^2+m^2)^{1/2}} \; .
\end{align}
Repeating the DHS analyticity arguments leads to the formula \cite{Castro:2017mfj}
\begin{align}\label{eq:DHSBTZ}
	Z_{\rm PI} = \prod_{k,l\in\mathbb{Z}} \prod_{z_l}\left( \omega_{|k|,l}+ i z_l\right)^{-1/2} \;,
\end{align}
where 
\begin{align}
     \omega_{k,l} \equiv \frac{2\pi k}{\beta}-il\Omega_H  \; , \qquad k,l\in\mathbb{Z}
\end{align}
are the Matsubara frequencies, and  we have emphasized the $l$-dependence of the QNM frequencies $z_l$. We have also assumed  PT-invariance so that for any (ingoing) QNM with frequency $z_l$, there is an (outgoing) anti-QNM with frequency $-z_{-l}$. Using $\log x = -\int_0^\infty \frac{dt}{t} e^{-x t}$ (ignoring the issue of UV-divergence), we can write \eqref{eq:DHSBTZ} in a more physically illuminating form \cite{Law:2022zdq}
\begin{align}\label{eq:spinningDHS}
	\log Z_{\rm PI} = \int_0^\infty \frac{dt}{2t} \coth \frac{\pi t}{\beta}\chi(t) \; , \qquad 	\chi(t ) = \sum_{l}\sum_{z_l} e^{i \left(l\Omega_H-z_l \right) t} \;. 
\end{align}
In this integral representation, the factor $\coth \frac{\pi t}{\beta}$ comes from the Matsubara sum over $k\in \mathbb{Z}$. The object $\chi(t)$ is the analog of the ``QNM character'' for static black holes introduced in \cite{Law:2022zdq}, with an extra twist $e^{i l \Omega_H t}$ by the black hole spin. This version of the DHS formula first appeared in the context of a static patch in de Sitter space \cite{Anninos:2020hfj} whose Euclidean counterpart is the round sphere; this was later extended to all static black holes in \cite{Law:2022zdq}, which established that the Fourier transform of the QNM character is a relative spectral density for the Hamiltonian for the free fields in the Lorentzian black hole geometry. 

The (ingoing) QNM spectrum for a scalar with mass $m^2=\Delta \left(\Delta -2 \right)$ is well-known \cite{Birmingham:2001pj}\footnote{The spectrum \eqref{eq:BTZQNM} is obtained using an ansatz of the form $\phi \propto e^{-i\omega t + i l \varphi}R(r)$.\label{footnote:convention}}
\begin{align}\label{eq:BTZQNM}
	\omega^L_{nl} = l -2\pi i T_L \left(2n+\Delta\right) \quad {\rm and} \quad \omega^R_{nl} = -l -2\pi i T_R \left(2n+\Delta\right) \; , \qquad n = 0,1,2,\cdots \;, \quad l \in \mathbb{Z} \; . 
\end{align}
With this we can compute the twisted character
\begin{align}\label{eq:btzchar}
	\chi(t) &= \sum_{n=0}^\infty \sum_{l\in\mathbb{Z}} \left(e^{i \left( l\Omega_H -\omega^L_{nl}\right) t }+e^{i \left( l\Omega_H -\omega^R_{nl}\right) t } \right) \nn\\
	&= \frac{e^{-2\pi T_L \Delta t}}{1-e^{-4\pi T_L t}} \sum_{l\in\mathbb{Z}}e^{-i l \frac{2\pi T_L }{r_+}t}+\frac{e^{-2\pi T_R \Delta t}}{1-e^{-4\pi T_R t}} \sum_{l\in\mathbb{Z}}e^{i l \frac{2\pi T_R}{r_+}t} \nn\\
	& = \frac{e^{-2\pi T_L \Delta t}}{1-e^{-4\pi T_L t}} \frac{r_+}{T_L}\sum_{p\in\mathbb{Z}}\delta \left(t-\frac{p r_+}{T_L} \right)  +\frac{e^{-2\pi T_R \Delta t}}{1-e^{-4\pi T_R t}} \frac{r_+}{T_R}\sum_{p\in\mathbb{Z}}\delta \left(t-\frac{p r_+}{T_R} \right)  \;. 
\end{align}
In the second line we have used
\begin{align}
	\Omega_H-1 = -\frac{2\pi T_L}{r_+}\; , \qquad \Omega_H+1 = \frac{2\pi T_R}{r_+} \; , 
\end{align}
and performed the $n$-sum. In the third line we have used Poisson summation
\begin{align}
	\sum_{l\in \mathbb{Z}} e^{\pm i 2\pi l \tau} = \sum_{p\in \mathbb{Z}} \delta \left(\tau-p\right) \; . 
\end{align}
Inserting \eqref{eq:btzchar} into the formula \eqref{eq:spinningDHS} and regularizing the $t$-integral by a cutoff $t=\epsilon>0$, the integral collapses 
\begin{align}\label{eq:scalarZ}
	\log Z_{\rm PI} = \sum_{p=1}^\infty\frac{1}{p} \frac{q_L^\frac{p\Delta}{2}}{1-q_L^p}\frac{q_R^\frac{p\Delta}{2}}{1-q_R^p} \; , \qquad q_{L/R}\equiv e^{-\left(2\pi \right)^2 T_{L/R} } \; . 
\end{align}
Finally, expanding $\frac{1}{1-x} =\sum_{n=0}^\infty x^n$ and summing over $p$, we arrive at
\begin{align}
	\log Z_{\rm PI} = -\sum_{n_L ,n_R =0}^\infty \log \left( 1- q_L^{\frac{\Delta}{2}+n_L}q_R^{\frac{\Delta}{2}+n_R}\right) 
\end{align}
which is the standard formula \eqref{eq:scalarDet} from the literature \cite{Giombi:2008vd,David:2009xg}
\begin{align}
	Z_{\rm PI} = \prod_{n_L ,n_R =0}^\infty \frac{1}{1- q_L^{\frac{\Delta}{2}+n_L}q_R^{\frac{\Delta}{2}+n_R}} \; . 
\end{align}

\subsection{Massive spin-$s$ }\label{sec:massHS}

A spin-$s$ field with mass $m^2 = \left( \Delta-s\right) \left(\Delta+s-2 \right) $ has the determinant
\begin{align}\label{eq:ZPImass}
	Z^{s, \Delta}_{\rm PI} = \det \left( -\nabla_s^2 + \Delta\left(\Delta-2 \right) -s\right)^{-1/2} \; . 
\end{align}
Here $-\nabla_s^2$ is the spin-$s$ transverse traceless Laplacian.

\subsubsection*{The bulk partition function}

Using the (ingoing) QNM spectrum \cite{Datta:2011za}
\begin{gather}\label{eq:massQNM}
	\omega^{\Delta, s , L,\mp}_{nl} = l -2\pi i T_L \left(2n+\Delta \mp s\right) \quad {\rm and} \quad \omega^{\Delta, s ,R,\mp}_{nl} = -l -2\pi i T_R \left(2n+\Delta\pm s\right) \; , \quad n = 0,1,2,\cdots \;, \quad l \in \mathbb{Z} \; , 
\end{gather}
we can compute the twisted QNM character just as in the scalar case, resulting in 
\begin{align}\label{eq:btzcharmass}
	\chi_{[\Delta, s]}(t) & = \frac{e^{-2\pi T_L \Delta t} \left( e^{2\pi T_L s t}+e^{-2\pi T_L s t}\right) }{1-e^{-4\pi T_L t}} \frac{r_+}{T_L}\sum_{p\in\mathbb{Z}}\delta \left(t-\frac{p r_+}{T_L} \right)  +\frac{e^{-2\pi T_R \Delta t}\left( e^{2\pi T_R s t}+e^{-2\pi T_R s t}\right) }{1-e^{-4\pi T_R t}} \frac{r_+}{T_R}\sum_{p\in\mathbb{Z}}\delta \left(t-\frac{p r_+}{T_R} \right)  \;.
\end{align}
Naively applying \eqref{eq:spinningDHS} then leads to 
\begin{align}\label{eq:masspinbulk}
	\log Z_{\rm bulk} = \sum_{p=1}^\infty\frac{1}{p} \frac{q_L^\frac{p\Delta}{2}}{1-q_L^p}\frac{q_R^\frac{p\Delta}{2}}{1-q_R^p}\left(q_L^{p\frac{s}{2}} q_R^{p\frac{s}{2}}+q_L^{-p\frac{s}{2}} q_R^{-p\frac{s}{2}}\right)  \; , \qquad q_{L/R}\equiv e^{-\left(2\pi \right)^2 T_{L/R} } \; . 
\end{align}

\subsubsection*{The edge partition function}

The result \eqref{eq:masspinbulk} is not the full answer. The formula \eqref{eq:DHSBTZ} assumes that for a fixed $k\in \mathbb{Z}$, any QNM with frequency $z(m^2)$ can Wick-rotate to a regular Euclidean mode with Matsubara number $|k|$ when $m$ hits the complex value $m_*$ such that $z(m_*^2) = i \omega_{|k|,l}$. This is true for scalars, but for spinning fields it fails for QNMs with frequencies 
\begin{align}\label{eq:btzedgefre}
	\omega^{\Delta, s , L,-}_{nl}= l - 2\pi T_L i \left( 2n+\Delta- s\right)\; ,    \qquad \omega^{\Delta, s , R,+}_{nl}=- l - 2\pi T_R i \left( 2n+\Delta - s\right)  \;, \quad n< s-|k| \; , \quad l\in \mathbb{Z} \; . 
\end{align}
This was first confirmed for $s\leq 2$ \cite{Castro:2017mfj,Keeler:2018lza} and later for any $s\geq 3$ \cite{Keeler:2019wsx,Grewal:2022hlo}. Because of this, these modes need to be excluded from the naive DHS formula for the corresponding $k$ sectors, leading to an ``edge" partition function: 
\begin{align}\label{eq:masspinedge}
	\log Z_{\rm edge} = &\int_0^\infty \frac{dt}{2t}\sum_{k=-(s-1)}^{s-1} \sum_{n=0}^{s-1-|k|} \sum_{l\in\mathbb{Z}}e^{-\frac{2\pi|k|}{\beta}t } \left(e^{i \left( l\Omega_H -\omega^{\Delta, s , L,-}_{nl}\right) t }+e^{i \left( l\Omega_H -\omega^{\Delta, s , R,+}_{nl}\right) t } \right)\nn\\
	=&\sum_{p=1}^\infty \frac{1}{p}  \frac{q_L^\frac{p\Delta}{2}}{1-q_L^p}\frac{q_R^\frac{p\Delta}{2}}{1-q_R^p} \left(q_L^{p\frac{s}{2}}-q_L^{-p\frac{s}{2}} \right) \left(q_R^{p\frac{s}{2}}-q_R^{-p\frac{s}{2}} \right)\; . 
\end{align}
Subtracting \eqref{eq:masspinedge} from \eqref{eq:masspinbulk} yields
\begin{align}
	\log Z_{\rm PI}^{ s, \Delta} =& \log Z_{\rm bulk}-\log Z_{\rm edge}  =\sum_{p=1}^\infty \frac{1}{p}  \frac{q_L^\frac{p\Delta}{2}}{1-q_L^p}\frac{q_R^\frac{p\Delta}{2}}{1-q_R^p} \left(q_L^{p\frac{s}{2}} q_R^{-p\frac{s}{2}}+q_L^{-p\frac{s}{2}} q_R^{p\frac{s}{2}}\right) \; . 
\end{align}
Finally, expanding $\frac{1}{1-x} =\sum_{n=0}^\infty x^n$ and summing over $p$, we arrive at
\begin{align}\label{eq:ZPImassresult}
	\log Z_{\rm PI}^{ s, \Delta} = -\sum_{n_L ,n_R =0}^\infty \log \left( 1- q_L^{\frac{\Delta+s}{2}+n_L}q_R^{\frac{\Delta-s}{2}+n_R}\right) \left( 1- q_L^{\frac{\Delta-s}{2}+n_L}q_R^{\frac{\Delta+s}{2}+n_R}\right) \;, 
\end{align}
which is the correct result \cite{David:2009xg}.

\subsection{Graviton}

The partition function for a graviton on Euclidean rotating BTZ is \cite{David:2009xg}
\begin{align} \label{eq:gravPI}
	Z_{\rm PI} = \frac{ \det \left( -\nabla_1^2 +2\right)^{1/2} }{ \det \left( -\nabla_2^2 -2\right)^{1/2} } \; . 
\end{align}
 This expression assumes that the contour for the conformal mode has been rotated so as to have a positive definite kinetic term. To compute the determinants involved in \eqref{eq:gravPI} we can use limits of the massive spin-$s$ results
\begin{align}\label{eq:gravlimits}
	s =1 : \qquad \Delta \to 3 \;, \qquad s=2: \qquad \Delta \to 2 \; . 
\end{align}
For our purposes it is more informative to recalculate the combined ratio of determinants in order to determine which graviton QNMs actually contribute to the answer.

\subsubsection*{The bulk partition function}

The bulk partition function is defined by the naive DHS formula, which in this case says
\begin{align}\label{eq:Zbulk}
	\log Z_{\rm bulk} = \int_0^\infty \frac{dt}{2t} \frac{1+e^{-2\pi t/\beta}}{1- e^{-2\pi t/\beta}}\left( \chi_{[\Delta=2 , s=2]}(t)- \chi_{[\Delta=3 , s=1]}(t)\right)  \; ,
\end{align}
where $\chi_{[\Delta, s]}(t) $ is defined in \eqref{eq:btzcharmass}. Now, it is easy to see that
\begin{align}
	\omega^{2, 2 , L,+}_{nl} = \omega^{3, 1 , L,+}_{nl} \; , \qquad \omega^{2, 2 , R,-}_{nl} = \omega^{3, 1 , R,-}_{nl} \; , \qquad \forall n=0 , 1, \cdots \; ,\quad l \in \mathbb{Z} 
\end{align}
and 
\begin{align}
	\omega^{2, 2 , L,-}_{n+1, l}= \omega^{3, 1 , L,-}_{n, l} \; , \qquad \omega^{2, 2 , R,+}_{n+1,l}= \omega^{3, 1 , R,+}_{n, l}\; , \qquad \forall n=0 , 1, \cdots \; ,\quad l \in \mathbb{Z} \; . 
\end{align}
Because of this there is a huge cancellation in  the bracket \eqref{eq:Zbulk}, which captures the physical QNMs. Explicitly,
\begin{align}\label{eq:cancellations}
	\chi_{[\Delta=2 , s=2]}(t)- \chi_{[\Delta=3 , s=1]}(t)= & \sum_{l\in\mathbb{Z}}\left(  e^{i \left( l\Omega_H -\omega^{2, 2 , L,-}_{n=0 , l}\right) t }+e^{i \left( l\Omega_H -\omega^{2, 2 , R,+}_{n =0, l}\right) t }\right) \nn\\
	=&  \frac{r_+}{T_L}\sum_{p\in\mathbb{Z}}\delta \left(t-\frac{p r_+}{T_L} \right) +  \frac{r_+}{T_R}\sum_{p\in\mathbb{Z}}\delta \left(t-\frac{p r_+}{T_R} \right) \; . 
\end{align}
Doing the $t$-integral and expanding the denominators gives 
\begin{align}\label{eq:Zgravbulk}
	\log Z_{\rm bulk} =& \sum_{p=1}^\infty\frac{1}{p}\frac{1- q_L^{p} q_R^{p} }{\left( 1-q_L^p\right) \left( 1-q_R^p\right) } \nn\\
	=& \sum_{n_L , n_R =0}^\infty \sum_{p=1}^\infty\frac{1}{p}\left(q_L^{n_L p} q_R^{n_R p}- q_L^{\left( n_L+1 \right) p} q_R^{\left( n_R+1\right) p}  \right)\nn\\
	=&  \sum_{p=1}^\infty\frac{1}{p}\left( \sum_{n_L=1}^\infty \sum_{n_R =1}^\infty q_L^{n_L p} q_R^{n_R p}+1 +\sum_{n_L=1}^\infty q_L^{n_L p}+ \sum_{n_R =1}^\infty  q_R^{n_R p}- \sum_{n_L , n_R =0}^\infty q_L^{\left( n_L+1 \right) p} q_R^{\left( n_R+1\right) p}  \right)\nn\\
	=&   \sum_{p=1}^\infty\frac{1}{p}\left( 1 +\sum_{n_L=1}^\infty q_L^{n_L p}+ \sum_{n_R =1}^\infty  q_R^{n_R p}  \right) \; . 
\end{align}
The first term is an infinite sum that cannot be regularized. It will be canceled by terms in the edge partition function.

\subsubsection*{The edge partition function}

Taking limits \eqref{eq:gravlimits} of the excluded modes \eqref{eq:btzedgefre}, we can write down the edge partition function
\begin{align}\label{eq:Zgravedge}
	\log Z_{\rm edge} = &\int_0^\infty \frac{dt}{2t}\sum_{k=-1}^{1} \sum_{n=0}^{1-|k|} \sum_{l\in\mathbb{Z}}e^{-\frac{2\pi|k|}{\beta}t } \left(e^{i \left( l\Omega_H -\omega^{2, 2 , L,-}_{nl}\right) t }+e^{i \left( l\Omega_H -\omega^{2, 2 , R,+}_{nl}\right) t } \right)\nn\\
	& - \int_0^\infty \frac{dt}{2t}  \sum_{l\in\mathbb{Z}} \left(e^{i \left( l\Omega_H -\omega^{3, 1 , L,-}_{n=0 , l}\right) t }+e^{i \left( l\Omega_H -\omega^{3, 1 , R,+}_{n=0,l}\right) t } \right)\nn\\
	= &\int_0^\infty \frac{dt}{2t}\sum_{k=-1}^{1} \sum_{l\in\mathbb{Z}}e^{-\frac{2\pi|k|}{\beta}t } \left(e^{i \left( l\Omega_H -\omega^{2, 2 , L,-}_{n=0, l}\right) t }+e^{i \left( l\Omega_H -\omega^{2, 2 , R,+}_{n=0, l}\right) t } \right)\nn\\
	=& \sum_{p=1}^\infty\frac{1}{p} \left( 1+ q_L^ p+ q_R^p\right) \; .
\end{align}
Taking the difference between \eqref{eq:Zgravbulk} and \eqref{eq:Zgravedge} results in the 1-loop part of \eqref{eq:AdSgravPI}. Notice that the shifts of $n_{L/R}\to n_{L/R}+2$ originate from the edge partition function \eqref{eq:Zgravedge}. 

 In appendix \ref{app:HS}, we generalize our considerations to massless higher spin fields in BTZ and obtain their low-temperature $\log T$ corrections.

\section{Which quasinormal modes produce $T^{3/2}$? }\label{sec:T32}
We now  turn to the question that motivated the recalculation of the near-extremal determinant. It is clear from \eqref{eq:cancellations} that almost none of the graviton QNMs contribute to the 1-loop determinant due to the massive cancellations with the isospectral ghost system. The two branches that contribute are the totally undamped modes
\begin{equation}
 \omega^{2, 2 , L,-}_{n=0 , l}=\ell \; , \qquad \qquad   \omega^{2, 2 , R,+}_{n =0, l}=-\ell \; 
\end{equation}
along with their anti-QNM counterparts. 
Of these two branches of modes, only one actually contributes to the low temperature scaling. According to \eqref{eq:cancellations} the left-moving undamped modes contribute a factor
\begin{equation}
\frac12 \sum_{p=1}^{\infty} \frac{1}{p}\frac{1+\exp[-\frac{2\pi p r_+}{\beta T_L}]}{1-\exp[-\frac{2\pi p r_+}{\beta T_L}]}= \frac12 \sum_{p=1}^\infty \frac{1}{p}\frac{1+q_R^p}{1-q_R^p} 
\end{equation}
while the right moving modes contribute
\begin{equation}\label{eq:RightMovers}
    \frac12 \sum_{p=1}^\infty \frac{1}{p}\frac{1+\exp[-\frac{2\pi p r_+}{\beta T_R}]}{1-\exp[-\frac{2\pi p r_+}{\beta T_R}]}=\frac12 \sum_{p=1}^\infty \frac{1}{p}\frac{1+q_L^p}{1-q_L^p} \;  =   \frac12 \sum_{p=1}^\infty \frac{1}{p}(1+2q_L^p)+\sum_{n=2}^\infty\sum_{p=1}^\infty\frac{1}{p}q_L^{np}. 
\end{equation}
Somewhat counter-intuitively, the left moving modes produce dependence on the right-moving temperature, and vice versa. 
Removing the contribution from the right moving edge modes
\begin{equation}
    \sum_{p=1}^\infty\frac{1}{p}\left(\frac12 + q_L^p\right) 
\end{equation}
yields the total right-moving contribution
\begin{equation}\label{eq:logZright}
    \log Z_{\text{right}}=-\sum_{n=2}^{\infty}\log(1-q_L^n)=\log \prod_{n=2}^\infty\frac{1}{1-q_L^n}\; . 
\end{equation}
Taking the small $T_L$ limit as in \eqref{eq:BTZprod} produces the $\frac32\log T$.

Since we know which modes account for the scaling, we can look for a shortcut in the product representation. 
The naive product form of the DHS formula using the undamped right moving modes is
\begin{align}
    (Z_{\text{right}})^2&=\prod_{k,l\in \mathbb{Z}}\frac{1}{2\pi|k|T-il\Omega_H -il}
    =\prod_{k,l\in \mathbb{Z}}\frac{1}{2\pi|k|\frac{2T_LT_R}{T_L+T_R}-\frac{2 ilT_R}{T_L+T_R}} \; .
\end{align}
According to \eqref{eq:Zgravedge} we are actually supposed to exclude $k=0,\pm 1$ from this product since those modes do not continue to regular solutions in Euclidean signature, so the correct formula is 
\begin{equation}\label{eq:Z32}
        Z_{\text{right}}=\prod_{k>1}\prod_{l\in \mathbb{Z}}\frac{1}{2\pi k\frac{2T_LT_R}{T_L+T_R}-\frac{2 ilT_R}{T_L+T_R}} \; . 
\end{equation}
In general, infinite constants of the form $\prod_{l\in\mathbb{Z}}\frac{1}{A}$
where $A$ is independent of $l$, can be absorbed into field redefinitions/local counterterms. For instance in $\zeta$-function regularization
\begin{align}
    \log \prod_{l\in\mathbb{Z}}\frac{1}{A} = -2 \zeta_R(0) \log A-\log A =0 \; . 
\end{align}
Here $\zeta_R(z)$ is the Riemann zeta function and  $\zeta_R(0)=-\frac12$. With this in mind, we have
\begin{equation}\label{eq:reducedProd}
        Z_{\text{right}}=\prod_{k>1}\prod_{l\in \mathbb{Z}}\frac{1}{2\pi k T_L- il}\; .
\end{equation}
Note that there is only dependence on $T_L$, as expected. Had we chosen the left-moving QNM, an overall factor of $T_L$ would have instead factored out and the product would only depend on $T_R$.
As a check of this formula, note that
\begin{equation}
    \log Z_{\text{right}}= -\sum_{k>1,l\in \mathbb{Z}}\log(2\pi k T_L -il)= \sum_{k>1,l\in \mathbb{Z}}\int \frac{dt}{t}e^{-(2\pi k T_L -il)t}=\sum_{p=1}^\infty\sum_{k=2}^\infty\frac{1}{p}e^{-(2\pi)^2T_L kp}=-\sum_{k=2}^\infty\log(1-q_L^k) 
\end{equation}
which is \eqref{eq:logZright}. 
The low-$T$ limit of this product therefore exhibits $T^{3/2}$ scaling. To see this explicitly we separate off the $l=0$ term so that the product becomes
\begin{equation}
        Z_{\text{right}}=\left[\prod_{k>1}\frac{1}{2\pi k T_L }\right]\left[\prod_{k>1}\prod_{l>0}\frac{1}{l^2 + (2\pi k T_L)^2}\right] \; . 
\end{equation}
The first factor is the familiar nearly-zero mode determinant from the throat calculation \eqref{eq:ZeroModeProd1}. One can explicitly check that the second factor does not introduce any overall factors of $T$. Heuristically this can be seen by taking the small-$T$ limits of the denominators in the second factor, which are nonsingular in the limit since $l>0$ (this argument is rigorous when we regulate the products). 

It will be interesting to try to find the factor \eqref{eq:reducedProd} in the Kerr DHS product.
It is not immediately clear that the analyticity assumptions that go into the derivation of the DHS formula \eqref{eq:DHSBTZ} will hold for asymptotically flat black holes. For instance, it is known that the analytic structure of the Fourier-domain retarded Green's functions for Kerr and Reissner-Nordstrom contain branch cuts along with the usual quasinormal poles. This reflects a physical effect: the time-domain response functions contain power-law tails (the Price tails \cite{Price:1971fb,Price:1972pw}) as well as exponentially damped contributions, reflecting the continual scattering of waves off of the long-range Coulombic potential of the black hole.\footnote{It is likely that superradiant effects and the lack of a true Hartle-Hawking state \cite{Kay:1988mu} will also complicate the analysis.}

Since this additional effect has more to do with the far region than the black hole itself, we are optimistic that it can be isolated and dealt with so that some form of the DHS argument still applies. In the next section we will provide
a new derivation of the DHS formula which we believe can be adapted to the asymptotically flat case.  Performing the analogous calculation in Kerr should should clarify any important differences with the asymptotically AdS case where the formula is known to hold. 

Assuming for the moment that the Kerr path integral can be accurately represented by a product of the form \eqref{eq:DHSBTZ}, we expect to encounter terms like

\begin{align}
    Z^2_{\rm PI} \supset \prod_{k}\prod_n\prod_{\ell,m} \frac{1}{\omega_{|k|,m}+i z_{n\ell m}}=\prod_{k}\prod_n\prod_{\ell,m} \frac{1}{2\pi |k| T - i m \Omega_H+i z_{n\ell m}} \; .
\end{align}
Here $n$ is the overtone number, $\ell$ is related to the separation constant in the polar direction and $m$ is the azimuthal number. To encounter power law dependence in $T$ in such an expression, we seem to need two things:
\begin{enumerate}
    \item For a finite (or a non-locally infinite) number of $(n,\ell,m)$, 
\begin{align}
    - i m \Omega_H+i z_{n\ell m} =0 \;.
\end{align}
This likely only occurs for massless fields and allows to pull out a product of $\Omega=0$ Matsubara frequencies.
    \item $k$ does not range over all integers. The exclusion of  QNMs with bad Euclidean continuations analogous to \eqref{eq:btzedgefre} will likely play a role here.
\end{enumerate}
If these conditions are met we might separate out a factor 
\begin{align}
    \prod_{k\in \mathbb{Z}\setminus \sigma} \frac{1}{2\pi |k| T}
\end{align}
(where $\sigma$ is some finite set of integers) which connects directly with the throat calculation.  Exploiting cancellations with the ghost system will also likely simplify the analysis.

\section{Eigenvalues and the BTZ  spectral measure}
\label{sec:BTZ}
For the reasons mentioned in the introduction, it is interesting to try to reproduce the result \eqref{eq:throatT32} using the full black hole geometry in as many ways as possible. We have already seen how to reproduce this scaling using the QNM spectrum of the black hole. In this section we reproduce the scaling using the eigenvalue definition of the determinant.

While the scaling \eqref{eq:throatT32} arises from a discrete set of nearly-zero modes in the not-NHEK throat, we find (perhaps surprisingly) that the factor of $T^{3/2}$ can be obtained in BTZ using only the essential (continuous) spectrum of the graviton kinetic operator.
In general it is difficult to determine whether or not a differential operator on a noncompact space of infinite volume has any normalizable eigenfunctions. In the process of discovering the zero modes of AdS$_2$, Camporesi and Higuchi \cite{Camporesi:1994ga} also proved that there is no point spectrum for the graviton in hyperbolic space $\mathbb{H}_n$ for $n>2$. Taking a discrete quotient can certainly introduce normalizable eigenfunctions, but \cite{Acosta:2021oqt} also claim that there is no point spectrum for gravitons on Euclidean BTZ.

In the sections that follow we explicitly solve for the essential spectrum of the BTZ kinetic operators and derive the corresponding spectral measure. The integral over these eigenvalues reproduces all known BTZ determinants without contributions from any point spectrum. We begin the section with a quick introduction to the relevant techniques. More details can be found in \cite{Camporesi:1994ga,BorthwickBook2016,baumgartel1983mathematical,dyatlov2019mathematical}. 

\subsection{Continuous spectra,  heat kernel, and the scattering phase}\label{sec:scattering}
When calculating the determinant of a positive elliptic differential operator $L$ (for example $-\nabla^2$) on a compact $d$-dimensional Euclidean manifold $M_d$ with a discrete spectrum $\Lambda_n$
\begin{equation}
    L\phi_n(x)=\Lambda_n \phi_n(x) \; , 
\end{equation}
it is often convenient to introduce an auxiliary object called the heat kernel $K(T,x,y)$ which solves the heat conduction problem on $\mathbb{R}\times M_d$:
\begin{equation}\label{eq:HeatEqn}
    \left(\partial_T + L_x\right)K(T,x,y)=0 \; , \qquad \qquad K(0,x,y)=\delta^{(d)}(x-y)\; .
\end{equation}
Since on a compact manifold the normalized eigenfunctions of $L$ provide a representation of the delta function ${\sum_n\phi_n(x)\phi^*_n(y)=\delta^{(d)}(x-y)}$, the heat kernel has a spectral representation
\begin{equation}
    K(T,x,y)=\sum_n\phi_n(x)\phi^*_n(y)e^{-\Lambda_n T} \; .
\end{equation}
Taking the trace of this equation
\begin{equation}\label{eq:heattrace}
    K(T)\equiv \int d^dx d^dy  \; K(T,x,y)\,\delta^{(d)}(x-y) = \sum_n e^{-\Lambda_n T}
\end{equation}
and using the identity $ \log \Lambda  =-\int_0^\infty \frac{ds}{s}e^{-\Lambda s}$ (ignoring the divergent piece)
one obtains an unregulated formal expression for the determinant of $L$:
\begin{equation}\label{eq:logdet}
    \log \det L = -\int_0^\infty \frac{dT}{T}K(T) \; .
\end{equation}
Since $K(T)=\text{Tr}\,e^{-LT}$ one could equivalently write
\begin{equation}\label{eq:discreteSpec}
    K(T)= \int_0^\infty d\Lambda\; \rho(\Lambda)\, e^{-\Lambda T} \; , \qquad \qquad \rho(\Lambda)=\sum_n \delta(\Lambda-\Lambda_n) \; .
\end{equation}
When the manifold $M_d$ is non-compact with infinite volume, the spectrum of $L$ is typically continuous and the definition of $K(T)$ itself requires IR-regulation. For noncompact Lie groups $G$ and their symmetric quotients $G/K$  there is an unambiguous ``unintegrated" heat trace that allows to give a formal definition to the determinant:
\begin{equation}\label{eq:ContHeatKernel}
    -\log \det L=``\text{vol}(G/K)"\int_0^\infty \frac{dT}{T} K(T,x_0,x_0)\; .
\end{equation}
In these cases there is typically a discrete degeneracy $\{k\}$ for each eigenvalue and the coincident heat kernel
\begin{equation}
    K(T,x_0,x_0)=\int d\Lambda \sum_{\{k\}}\phi^*_{\Lambda, \{k\}}(x_0)\phi_{\Lambda, \{k\}}(x_0)e^{-\Lambda T}
\end{equation}
is actually independent of its coordinate argument due to the symmetries of the problem. Taking the integral over $K(T,x_0,x_0)$ produces a factor of the regularized volume multiplied by an integral over the spectral measure
\begin{equation}
 \rho(\Lambda)=\sum_{\{k\}}\phi^*_{\Lambda, \{k\}}(x_0)\phi_{\Lambda, \{k\}}(x_0)
\end{equation}
which can be evaluated at any point. This is one way to evaluate the determinant on $\mathbb{H}_3$, which simply renormalizes the cosmological constant due to the volume factor in \eqref{eq:ContHeatKernel}.

When there are not enough symmetries to fix the coordinate dependence of the heat kernel, there are different ways to proceed. Rather than a discrete sum like \eqref{eq:heattrace}, one expects to encounter an integral
\begin{equation}\label{eq:IntLambda}
    \log \det L=-\int_0^\infty \frac{dT}{T}\int_{\Lambda_0}^\infty d\Lambda \, \rho(\Lambda)e^{-\Lambda T} \;.
\end{equation}
The spectral measure $\rho(\Lambda)$ is a continuous generalization of \eqref{eq:discreteSpec} for eigenfunctions $\phi_\Lambda(x)$ which are delta-normalizable
\begin{equation}
    \int d^dx \, \phi_{\Lambda}^*(x)\phi_{\Lambda'}(x)=\delta(\Lambda-\Lambda') \; .
\end{equation}
It is often easier to calculate $K(T)$ directly using the definition \eqref{eq:HeatEqn} rather than the integral \eqref{eq:IntLambda}. This is particularly true for discrete quotients like $\Gamma\backslash M$ when the heat kernel on $M$ is exactly solvable. Examples include thermal AdS$_3$ and the BTZ black hole, which are both discrete quotients of hyperbolic space $\Gamma\backslash \mathbb{H}_3$. When $\Gamma$ is cyclic and generated by $\gamma$ the method of images produces a heat kernel on the quotient that automatically satisfies \eqref{eq:HeatEqn}:
\begin{equation}
K(T,x,y)_{\Gamma\backslash\mathbb{H}_3}=\sum_n K(T,x,\gamma^n y)_{\mathbb{H}_3} \; .
\end{equation}
This quantity can then be integrated and regulated to produce the determinant (the dependence on $(x,y)$ is typically nontrivial). This is the way that the BTZ path integral has traditionally been calculated \cite{Giombi:2008vd} and matched to CFT$_2$ characters. Although this technique is powerful, it sidesteps the direct description of the spectrum of the differential operator and makes it difficult to compare to the calculation performed in section \ref{sec:throat}. For that reason we would like to determine $\rho(\Lambda)$ explicitly and do the integral to reproduce the known results and compare to the throat calculation. In particular, we would like to know how the $T^{3/2}$ scaling is encoded in the spectral measure $\rho(\Lambda)$.

Another motivation to analyze the spectral measure is that generic manifolds including Kerr are not (quotients of) homogeneous spaces, so the discussions above simply do not apply. Fortunately, there is an alternative mathematically well-defined way to make sense of the spectral measure (see for instance \cite{baumgartel1983mathematical}). For concreteness, we consider 
the eigenvalue problem for the Laplacian on Euclidean black holes (including BTZ and Kerr) where the eigenvalue PDE
\begin{align}\label{eq:eigenPDE}
    -\nabla^2 \Psi_{\Lambda} = \Lambda \Psi_{\Lambda}
\end{align}
can be separated. In this case, we can separate out a radial function $R_{\Lambda \sigma}(r)$ ($\sigma$ collectively denotes other labels such as angular momentum) that obeys a 2nd-order ODE, which can be transformed into the normal form 
\begin{equation}\label{eq:RadialEq}
    \left(-\partial_r^2+ V_{\Lambda \sigma} (r)\right) \bar R_{\Lambda \sigma}(r)=0 \; , \qquad r_0\leq r<\infty \; . 
\end{equation}
The radial coordinate $r$ is chosen so that the boundary is at $r\to\infty$. We will see in the Euclidean BTZ example that the eigenvalue is more naturally parametrized by a spectral parameter $\lambda\in \mathbb{R}$, i.e. $\Lambda = \Lambda (\lambda)$, in terms of which the potential in \eqref{eq:RadialEq} takes the form 
\begin{align}
    V_{\Lambda \sigma}(r) \sim -\lambda^2 \qquad {\rm as } \quad r \to \infty \; . 
\end{align}
Because of this asymptotic form, the radial function $\bar R_{\lambda \sigma}(r)$ is a linear combination of plane waves at $r\to\infty$
\begin{equation}
    \bar R_{\lambda \sigma}(r\to \infty)\sim C_\sigma^{\text{in}}(\lambda) \, e^{- i \lambda r}+ C_\sigma^{\text{out}}(\lambda)\, e^{i \lambda r} \; .
\end{equation} 
Therefore, viewing \eqref{eq:RadialEq} as a stationary scattering problem, we can think of an eigenfunction as a scattering wavefunction; the ratio of the incoming and outgoing coefficients defines a reflection coefficient, or scattering phase
\begin{equation}\label{eq:scatteringphase}
    \theta(\lambda)=\frac{1}{2i}\log \frac{C^{\text{out}}(\lambda)}{C^{\text{in}}(\lambda)} \; .
\end{equation}
To define a spectral density, we first put an IR cutoff at $r=K$ so that the spectrum becomes discrete, i.e. $\lambda = \lambda_n$ are now labeled by integers $n=1,2,\cdots$. Solving for $\lambda_n$ in the large-$K$ limit, one can show that the density in $\lambda$ takes the form \cite{Law:2022zdq}
\begin{equation}
    \rho (\lambda) = \frac{1}{\Delta \lambda} \sim \frac{K+\theta'(\lambda)}{\pi } + O(K^{-1}) \; .
\end{equation}
The leading Weyl term $\frac{K}{\pi }$ indicates that in the unregulated problem, there are infinitely many eigenvalues in any interval within the spectrum in the continuum limit $K\to \infty$. The subleading term with the scattering phase $\theta(\lambda)$ depends on the exact form of the potential \eqref{eq:RadialEq} and contains the non-trivial information about the spectrum. The strategy is to choose some reference system (i.e. a stationary scattering problem \eqref{eq:RadialEq} with a different potential), and define a relative spectral density by taking the difference (so that we can safely send $K\to \infty$)
\begin{equation}\label{eq:renspecden}
 \mu(\lambda)\equiv  \lim_{K\to \infty} \left( \rho (\lambda)-\rho_0(\lambda) \right)= \frac{\theta'(\lambda)-\theta_0'(\lambda)}{\pi }\;.  
\end{equation}
This is sometimes known as the Krein-Friedel-Lloyd formula. It is worth noting that the scattering perspective for Euclidean BTZ has been discussed in \cite{perry2003selberg}.\footnote{See for instance \cite{diaz2009holographic,aros2010functional} for generalizations to higher dimensions.} However, as far as we know, the spectral measure has not been worked out through the relation \eqref{eq:renspecden}. We fill this gap in this section, and reproduce the results of \cite{Giombi:2008vd,David:2009xg} using a continuous integral without the inclusion of any discrete spectrum.

On compact manifolds the eigenvalue spectrum encodes a wealth of information about the underlying geometry. When the spectrum is continuous, the information about the geometry must instead be encoded in the generalized density of states $\mu(\lambda)$. Indeed, although $\mu(\lambda)$ is naively a function of a real argument, it can often be continued into the complex plane where it  exhibits poles related to resonant frequencies of the Wick-rotated operator. Deforming the contour of integration off of the real axis and using the residue theorem then allows one to convert the continuous integral over eigenvalues into a discrete sum over complex resonances, which turns out to be the DHS representation of the logarithm of the determinant. This discrete sum can then be used to easily make contact with \eqref{eq:ZeroModeProd1} as in section \ref{sec:DHS}.

The simplest nontrivial example that illustrates all of these ideas is the free scalar field on Euclidean BTZ, which we treat in section \ref{sec:ScalarSpec}. We use this example to illustrate the relationship between the spectral measure and the DHS representation of the determinant in section \ref{sec:dhsFromMu}, which includes a new proof of the DHS formula. Section \ref{sec:spinspec} derives the spin-$s$ spectral measure, expresses it in terms of the scalar spectral measure, and explains how the bulk/edge split encountered in section \ref{sec:DHS} arises in the eigenvalue computation.

\subsection{Scalar field spectral measure} \label{sec:ScalarSpec}

The spectral measure comes from solving the eigenvalue problem directly on the BTZ background. As before we set the AdS length $\ell=1$.  It is convenient to introduce the coordinate \cite{Birmingham:2001pj}
\begin{align}
    \tanh^2\xi = \frac{r^2-r^2_+}{r^2-r^2_-} 
\end{align}
so that the Lorentzian metric \eqref{eq:BTZBLmetric} takes the simple form
\begin{align}\label{eq:EBTZregmertric}
		ds^2=  d\xi^2 - \sinh^2\xi (r_+dt-r_-d \varphi)^2+   \cosh^2 \xi (r_-dt-r_+d\varphi)^2   \; , \qquad \varphi \sim \varphi + 2\pi \; . 
\end{align}
The conformal boundary is at $\xi \to \infty$. Wick rotating $t=-i\tau $ and changing the coordinate $z=\tanh^2\xi$, we look for eigenfunctions $-\nabla^2\Phi_\lambda =(1+\lambda^2)\Phi_\lambda$ using the ansatz
\begin{equation}\label{eq:RotatingMat}
    \Phi_{kl\lambda}(\tau, z, \varphi)=e^{-i\omega_{kl} \tau}e^{il\varphi}R_{kl\lambda}(z)\; , \qquad \qquad \omega_{kl}=\frac{2\pi k}{\beta}-il\Omega_H  \; , \qquad k,l\in\mathbb{Z} \; . 
\end{equation}
The spectrum of Matsubara frequencies arises from the identification ${(\tau,\varphi)\to (\tau+\beta, \varphi-i\beta\Omega_H)}$ required for a smooth Euclidean section.\footnote{In this section $\Omega_H$ is purely imaginary so that the Euclidean metric and the coordinate identifications ${(\tau,\varphi)\to (\tau+\beta, \varphi-i\beta\Omega_H)}$ are real.}
The radial eigenfunctions satisfy \cite{Birmingham:2001pj}
\begin{align}\label{eq:scalarSpec}
 z(1-z)\partial_z^2R_{kl\lambda} +(1-z)\partial_z R_{kl\lambda}(z) + \left[\frac{k_+^2}{4z}-\frac{k_-^2}{4}+\frac{1+\lambda^2}{4(1-z)}\right]R_{kl\lambda}(z)=0 \; ,  
\end{align}
where we have defined
\begin{equation}
    k_+=\frac{i\omega_{kl}r_+-lr_-}{r_+^2-r_-^2} \; ,
    \qquad k_-=\frac{i\omega_{kl} r_--lr_+}{r_+^2-r_-^2} \; . 
\end{equation}
Note that due to the form of the Matsubara frequencies \eqref{eq:RotatingMat}, along with $2\pi r_+T=r_+^2-r_-^2$ and $\Omega_H=r_-/r_+$, we have $  k_+=ik$.
Making the ansatz $R_{kl\lambda}(z)=z^\alpha(1-z)^\beta F_{kl\lambda}(z)$ turns \eqref{eq:scalarSpec} into a hypergeometric equation for $F_{kl\lambda}(z)$ for any of the four combinations $\alpha =\pm ik_+/2$ and $\beta=\frac12(1\pm i\lambda)$. Choosing $\alpha=-ik_+/2$ and $\beta=\frac12(1+i\lambda)$ we find 
\begin{equation}
    z(1-z)\partial_z^2 F +\left(1-ik_+ -z(2+i\lambda -ik_+)\right)\partial_z F- \frac14\left(k_+^2 +(1+i\lambda -ik_+)^2\right)F=0 \; ,
\end{equation}
whose two solutions are
\begin{equation}
    F^+_{kl\lambda}(z)= \, _2F_1\left(\frac{1+i\lambda}{2} -\frac{i}{2}(k_++k_-)\; ;\; \frac{1+i\lambda}{2} -\frac{i}{2}(k_+-k_-)\; ;\;1-ik_+ \; ;\; z\right) 
\end{equation}
and
\begin{equation}
    F_{kl\lambda}^-(z)= z^{ik_+}\, _2F_1\left(\frac{1+i\lambda}{2} +\frac{i}{2}(k_+-k_-)\; ;\; \frac{1+i\lambda}{2} +\frac{i}{2}(k_++k_-)\; ;\;1+ik_+ \; ;\; z\right)\; .
\end{equation}
In order to be a permissible generalized eigenfunction $R_{kl\lambda}^\pm$ needs to be regular at the origin ($z=0$), and this depends on the sign of $k$. Since $z^{-ik_+/2}=z^{k/2}$, $R^+$ is regular when $k>0$ and $R^-$ is regular when $k<0$.
In terms of the original variables
\begin{equation}\label{eq:PositiveKscalar}
 R_{kl\lambda}^+(\xi)=   (\tanh\xi)^{k}(\text{sech}\xi)^{1+i\lambda}\;_2F_1\left(\frac{1+i\lambda}{2}-\frac{i}{2}(ik+k_-);\frac{1+i\lambda}{2} -\frac{i}{2}(ik-k_-) ;1+k;\tanh^2\xi\right) \; . 
\end{equation}
Meanwhile, when $k<0$ the solution $R^-_{kl}(z)$ takes the form
\begin{equation}
    R_{kl\lambda}^-(z)=(\tanh\xi)^{|k|}(\text{sech}\xi)^{1+i\lambda}\;_2F_1\left(\frac{1+i\lambda}{2} -\frac{i}{2}(i|k|+k_-);\frac{1+i\lambda}{2}-\frac{i}{2}(i|k|-k_-) ;1+|k|;\tanh^2\xi\right) \; , 
\end{equation}
so that for any $k$ we can write
\begin{equation}
    R_{kl\lambda}(z)=(\tanh\xi)^{|k|}(\text{sech}\xi)^{1+i\lambda}\;_2F_1\left(\frac{1+i\lambda}{2} -\frac{i}{2}(i|k|+k_-);\frac{1+i\lambda}{2}-\frac{i}{2}(i|k|-k_-) ;1+|k|;\tanh^2\xi\right) \; . 
\end{equation}
In order to determine the spectral measure we need to determine the asymptotics of these solutions near the boundary at $z=1$.
Using the connection formula
\begin{align}\label{eq:2f1 form}
	_2F_1 (a,b;c;z ) =&\frac{\Gamma(c)\Gamma(c-a-b)}{\Gamma(c-a)\Gamma(c-b)}\,_2F_1 \left(a,b;a+b-c+1;1-z \right)\nn\\
	&+\frac{\Gamma(c)\Gamma(a+b-c)}{\Gamma(a)\Gamma(b)}(1-z)^{c-a-b}\,_2F_1 \left(c-a,c-b;c-a-b+1;1-z \right) 
\end{align}
as well as $\tanh \xi \to 1$, $\text{sech } \xi \to 2\, e^{-\xi}$ we have that\footnote{To compare with section \ref{sec:scattering} more directly, one can rescale $R_{kl\lambda}=\text{csch}^{\frac12}(2\xi)\bar R_{kl\lambda}$ so that $\bar R_{kl\lambda}$ obeys an ODE in the normal form \eqref{eq:RadialEq}.}
\begin{equation}
    R_{kl\lambda}\sim C_{kl}(\lambda)\, e^{-\xi}e^{-i \lambda \xi} + C_{kl}(-\lambda) \, e^{-\xi}e^{i\lambda \xi}\; . 
\end{equation}
It is clear that $\lambda \in \mathbb{R}$ if the eigenfunction is to be delta function normalizable. Moreover, we see that sending $\lambda \to -\lambda$ yields the same solution. The ingoing coefficient takes the factorized form
 \begin{gather}\label{eq:cklnew}
    C_{kl}(\lambda)=C^{\rm BTZ}_{kl}(\lambda)C^{\rm Rin}_{kl}(\lambda)\; , \\
    C^{\rm BTZ}_{kl}(\lambda)=\frac{\Gamma(1+|k|)}{\Gamma\left(\frac{1-i\lambda}{2}-\frac{i}{2}(i|k|-k_-)\right)\Gamma\left(\frac{1-i\lambda}{2} -\frac{i}{2}(i|k|+k_-)\right)}\; , \qquad C^{\rm Rin}_{kl}(\lambda)=2^{1+i\lambda}\Gamma(-i\lambda)\; .
\end{gather}
 We have separated out a $(k,l)$-independent factor $C^{\rm Rin}_{kl}(\lambda)$ that only depends on the asymptotics of the Euclidean BTZ space \cite{BorthwickBook2016}, and which can be thought of as the ingoing coefficient for the wave equation on Rindler space \cite{Law:2022zdq} at inverse temperature $2\pi$. As explained in the beginning of this section, we can define a renormalized spectral density through \eqref{eq:renspecden} with an appropriate reference subtraction. It turns out that the correct renormalized density of states is given by subtracting off the Rindler-like contribution
\begin{align}
    \mu_{kl}(\lambda)&=\frac{1}{2\pi i}\partial_\lambda \log\frac{C_{kl}(-\lambda)}{C_{kl}(\lambda)} -\frac{1}{2\pi i}\partial_\lambda \log\frac{C^{\rm Rin}_{kl}(-\lambda)}{C^{\rm Rin}_{kl}(\lambda)} = \frac{1}{2\pi i}\partial_\lambda \log \frac{C_{kl}^{\rm BTZ}(-\lambda)}{C_{kl}^{\rm BTZ}(\lambda)} \;   \nn\\
    &= \frac{1}{2\pi i}\partial_\lambda \log \frac{\Gamma\left(\frac{1-i\lambda}{2}-\frac{i}{2}(i|k|-k_-)\right)\Gamma\left(\frac{1-i\lambda}{2} -\frac{i}{2}(i|k|+k_-)\right)}{\Gamma\left(\frac{1+i\lambda}{2}-\frac{i}{2}(i|k|-k_-)\right)\Gamma\left(\frac{1+i\lambda}{2} -\frac{i}{2}(i|k|+k_-)\right)} \; . 
\end{align}
Using the series formula for the digamma function
\begin{equation}
    \psi(x)\equiv\frac{\Gamma'(x)}{\Gamma(x)}=-\gamma + \sum_{n=0}^\infty \left(\frac{1}{n+1}-\frac{1}{n+x}\right) 
\end{equation}
and performing another subtraction of the infinite but $(k,l)$-independent sum and $\gamma$ term we find 
\begin{align}\label{eq:mukl}
    \mu_{kl}(\lambda)&=\frac{1}{2\pi i}\sum_{n=0}^\infty\left(\frac{1}{\lambda -i(2n+1+|k|+ik_-)}-\frac{1}{\lambda + i(2n+1+|k|+ik_-)}\right) \nn \\
    &+ \frac{1}{2\pi i}\sum_{n=0}^\infty\left(\frac{1}{\lambda -i(2n+1+|k|-ik_-)}-\frac{1}{\lambda + i(2n+1+|k|-ik_-)}\right) \; \nn  \\
    &\equiv \frac{1}{2\pi i}\sum_{\pm}\sum_{n=0}^\infty\left(\frac{1}{\lambda +\lambda^\pm_{nkl}}-\frac{1}{\lambda -\lambda^\pm_{nkl}}\right) \; , 
\end{align}
where we have defined
\begin{equation}
    \lambda_{nkl}^{\pm}= -i(2n+1+|k|\pm ik_-) \; . 
\end{equation}
The poles in this expression correspond to the resonant frequencies of Euclidean rotating BTZ. 
Defining 
\begin{equation}
    z_{nl}^+ = -l -2\pi i T_R(2n+1) \; , \qquad z_{nl}^-=l-2\pi i T_L(2n+1) \; , 
\end{equation}
as in \eqref{eq:BTZQNM} and using 
\begin{equation}
    k_- = i k \frac{r_-}{r_+}-\frac{l}{r_+}\; ,  \qquad \frac{1}{T} = \frac{r_+}{2\pi T_L T_R} \; , 
    \qquad \frac{1}{r_+}=\frac{1\pm \Omega_H}{2\pi T_{R/L}} \; , 
\end{equation}
we have
\begin{align}\label{eq:lambdas}
    i\lambda_{nkl}^{\pm}= 2n+1+|k|\pm i \left(i k \frac{r_-}{r_+}-\frac{l}{r_+}\right) = 2n+1+|k|\mp  k \frac{r_-}{r_+}\mp i \frac{l}{r_+} \; . 
\end{align}
When $k$ is positive, we can write
\begin{align}
    i\lambda_{nkl}^{\pm}=   2n+1+\frac{2\pi kT}{2\pi T_{R/L}}- \frac{ i l \Omega_H \pm i l}{2\pi T_{R/L}}
    = \frac{\omega_{kl}+iz_{n,l}^{\pm}}{2\pi T_{R/L}} \qquad \text{for \;\;} k>0 \; .  
\end{align}
Similarly, when $k$ is negative we have
\begin{align}
    i\lambda_{nkl}^{\pm}=   2n+1-\frac{2\pi kT}{2\pi T_{L/R}}- \frac{ - i l \Omega_H \pm i l}{2\pi T_{L/R}} 
    = \frac{\omega_{-k,-l}+iz_{n,-l}^{\mp}}{2\pi T_{L/R}} \qquad \text{for \;\;} k<0 \; . 
\end{align}
So in summary
\begin{equation}\label{eq:scalarPoles}
    i\lambda_{nkl}^{\pm}=\frac{\omega_{kl}+iz_{n,l}^{\pm}}{2\pi T_{R/L}} \qquad \text{for \;\;} k>0\; , \qquad \qquad 
    i\lambda_{nkl}^{\pm}=\frac{\omega_{-k,-l}+iz_{n,-l}^{\mp}}{2\pi T_{L/R}} \qquad \text{for\;\;} k<0 \; . 
\end{equation}
In order to determine the spectral density at an eigenvalue $\lambda$ we sum over the generalized degeneracies $\mu_{kl}(\lambda)$
\begin{equation}
    \mu(\lambda)=\sum_{k,l\in \mathbb{Z}}\mu_{kl}(\lambda) \; .
\end{equation}
Since the imaginary parts of the $\lambda_{nkl}^{\pm}$ are negative we can write
\begin{equation}\label{eq:SigmaInt}
    \mu(\lambda)= \frac{1}{2\pi i }\sum_{k,l\in \mathbb{Z}}\sum_{\pm}\sum_{n=0}^\infty \left(\frac{1}{\lambda+\lambda_{nkl}^\pm}-\frac{1}{\lambda-\lambda_{nlk}^\pm}\right) 
    = \int_0^\infty \frac{dt}{2\pi}\left(e^{-i\lambda t}+e^{i\lambda t} \right)\sum_{k,l\in \mathbb{Z}}\sum_{\pm}\sum_{n=0}^\infty e^{-i\lambda_{nkl}^\pm t} \; . 
\end{equation}
Performing the sum over $n$ and using Poisson summation $\sum_{l\in \mathbb{Z}} e^{\pm i 2\pi l \tau} = \sum_{p\in \mathbb{Z}} \delta \left(\tau-p\right)$ we find
\begin{align}
    \sum_{k,l\in \mathbb{Z}} \sum_{\pm}\sum_{n=0}^\infty e^{-i\lambda_{nkl}^\pm t}  
    & =\frac{e^{-t}}{1-e^{-2t}}\sum_\pm \sum_{k,l\in \mathbb{Z}}e^{- \left( |k|\mp  k \frac{r_-}{r_+}\mp i \frac{l}{r_+} \right)t }\nn \\
    &=\frac{1}{2\sinh t}\sum_{l\in\mathbb{Z}}e^{ i\frac{l}{r_+}t} \sum_{k\in \mathbb{Z}}\left(e^{- \left( |k|-  k \frac{r_-}{r_+} \right)t }+e^{- \left( |k|+  k \frac{r_-}{r_+} \right)t }\right) \nn\\
    &=\frac{2\pi r_+}{2\sinh 2\pi r_+ p}\sum_p\delta(t-2\pi r_+ p)  \sum_{k\in \mathbb{Z}}\left(e^{- |k|\left( 1-  \Omega_H\right)t }+e^{- |k|\left( 1+\Omega_H  \right)t }\right)\; . 
\end{align}
In the last step we exchanged the $k<0$ terms in the two factors. Performing the geometric sums gives
\begin{equation}
    \sum_{k,l\in \mathbb{Z}} \sum_{\pm}\sum_{n=0}^\infty e^{-i\lambda_{nkl}^\pm t} 
    =\sum_p \delta (t-2\pi r_+ p)\frac{2\pi r_+}{2\sinh 2\pi p r_+}\left[\frac{1+e^{-(2\pi)^2T_Lp}}{1-e^{-(2\pi)^2T_Lp}}+\frac{1+e^{-(2\pi)^2T_Rp}}{1-e^{-(2\pi)^2T_Rp}}\right] \;  
\end{equation}
on the support of the delta function.
Integrating and discarding the infinite $p=0$ term  gives the spectral measure
\begin{align}\label{eq:ScalarMeasure}
\mu (\lambda)= 	  \sum_{p=1}^\infty\frac{r_+}{\sinh \pi p\frac{2\pi}{\beta_L}\sinh \pi p\frac{2\pi}{\beta_R}}  \cos 2\pi pr_+ \lambda  \; . 
\end{align}
Now we turn to the heat kernel 
\begin{align}
    K(T) \equiv \text{Tr} \, e^{-\left(-\nabla^2 \right) T}=  \frac12 \int_\mathbb{R} d\lambda \, \mu(\lambda) \, e^{-(1+\lambda^2)T}\; , 
\end{align}
where we insert the factor $\frac12$ since $\Phi_\lambda = \Phi_{-\lambda}$.  We compute the integral using
\begin{equation}\label{eq:Tintegralformula}
    \frac12\int_{-\infty}^\infty \cos(2\pi p \lambda \tau_2) \, e^{-(1+\lambda^2)T}d\lambda=\frac12\sqrt{\frac{\pi}{T}}\, e^{-T-(\pi p \tau_2)^2/T} \;.  
\end{equation}
Recalling that for BTZ the modular parameters are
\begin{equation}
    \tau=\frac{2\pi i}{\beta_L} \; , \qquad \bar{\tau}=-\frac{2\pi i}{\beta_R} \; , \qquad \tau_2=r_+ \; , 
\end{equation}
one finds the result commonly obtained using the method of images \cite{Giombi:2008vd} 
\begin{equation}\label{eq:s0heatkernel}
    K(T)=\sqrt{\pi}\sum_{p=1}\frac{ \tau_2}{|\sin p\pi \tau|^2}\frac{e^{-T-(\pi p \tau_2)^2/T}}{2\sqrt{T}} \; . 
\end{equation}
 Using the integral
\begin{equation}
    \sqrt{\pi}\int_0^\infty\frac{dT}{T}\frac{e^{-T-(\pi p \tau_2)^2/T}}{\sqrt{T}}=\frac{1}{p\tau_2}e^{-2\pi p \tau_2} 
\end{equation}
we then reproduce \eqref{eq:scalarZ}  with $\Delta =2$
\begin{equation}
    \log Z = \frac14 \sum_{n=1}^\infty\frac{1}{n}e^{-2\pi n \tau_2}\frac{ 1}{|\sin n\pi \tau|^2}=\sum_{n=1}^\infty\frac{1}{n}\frac{(q_Lq_R)^n}{(1-q_L^n)(1-q_R^n)} \; . 
\end{equation}

\subsection{DHS formula from the spectral measure }\label{sec:dhsFromMu}
Now we would like to use the spectral representation of the determinant to re-derive the DHS formula.  The path integral for a scalar with mass $m^2=\Delta (\Delta-2)$ has the heat kernel representation 
\begin{align}\label{eq:BTZPI}
	\log Z_{\rm PI} =& \int_0^\infty \frac{dT}{2T}  e^{-m^2 T}	K(T)\; , \qquad  e^{-m^2 T}K(T) 
 =\frac12 \int_{-\infty}^\infty d\lambda \,  \mu(\lambda) \, e^{-\left( \lambda^2+\left(\Delta-1 \right)^2  \right) T} \; . 
\end{align}
 Instead of substituting \eqref{eq:s0heatkernel} and computing the $T$-integral, we use the representation \eqref{eq:SigmaInt} for the spectral measure. One can first compute the $\lambda$- and $T$-integrals, leading to  
\begin{align}\label{eq:specDHS}
	\log Z_{\rm PI} &=\int_0^\infty \frac{dt}{2t} \, e^{-\left( \Delta -1\right)t }\sum_{k,l\in\mathbb{Z}}\sum_\pm \sum_{n=0}^\infty e^{-i \lambda^\pm_{nkl}t} \nn\\
 &=\int_0^\infty \frac{dt}{2t} \, \sum_{k,l\in\mathbb{Z}} \sum_{n=0}^\infty \left( e^{\left(-i\lambda^+_{nkl}- \Delta +1 \right)t} +e^{\left(-i\lambda^-_{nkl}- \Delta +1 \right)t}\right) \;.  
\end{align}
With \eqref{eq:scalarPoles} we can compute 
\begin{align}\label{eq:resonancecomptue}
    -i\lambda_{nkl}^{\pm}- \Delta +1&=\frac{-\omega_{kl}-iz_{n,l}^{\pm}- (\Delta -1)2\pi T_{R/L}}{2\pi T_{R/L}}=-\frac{\omega_{kl}+ i\omega_{n,l}^{R/L}}{2\pi T_{R/L}} \qquad \text{for \;\;} k>0\; ,  \nn\\
    -i\lambda_{nkl}^{\pm}- \Delta +1&=\frac{-\omega_{-k,-l}-iz_{n,-l}^{\mp}- (\Delta -1)2\pi T_{L/R}}{2\pi T_{L/R}}=-\frac{\omega_{-k,-l}+ i\omega_{n,-l}^{L/R}}{2\pi T_{L/R}} \qquad \text{for\;\;} k<0 \; ,
\end{align}
where we recall the Lorentzian QNM frequencies  \eqref{eq:BTZQNM}.   
Substituting these back in \eqref{eq:specDHS} and rescaling away the denominators in \eqref{eq:resonancecomptue} by taking $t \to 2\pi T_{L/R}t $, we have
\begin{align}
	\log Z_{\rm PI} 
 &=\int_0^\infty \frac{d t}{2t} \, \sum_{l\in\mathbb{Z}} \sum_{n=0}^\infty \left[ \sum_{k\geq 0}\left( e^{-\left(\omega_{kl}+ i\omega_{n,l}^{R} \right)t} + e^{-\left(\omega_{kl}+ i\omega_{n,l}^{L} \right)t} \right) +\sum_{k<0}\left( e^{-\left(\omega_{-k,-l}+ i\omega_{n,-l}^{L}\right)t} + e^{-\left(\omega_{-k,-l}+ i\omega_{n,-l}^{R} \right)t} \right)  \right]\nn\\
 &=\int_0^\infty \frac{dt}{2t} \, \sum_{l\in\mathbb{Z}} \sum_{n=0}^\infty  \sum_{k\in \mathbb{Z}}\left( e^{-\left(\omega_{|k|l}+ i\omega_{n,l}^{R} \right)t} + e^{-\left(\omega_{|k|l}+ i\omega_{n,l}^{L} \right)t} \right) \;, 
\end{align}
which is exactly the DHS form \eqref{eq:DHSBTZ} of the scalar partition function.  Now, we see that each factor in the DHS formula \eqref{eq:DHSBTZ} corresponds to a resonance pole in the spectral measure \eqref{eq:mukl}.

\subsection{Spin-$s$ spectral measure and determinant}\label{sec:spinspec}

A massive spin-$s$ field is described by the Lorentzian Fierz-Pauli system
\begin{align}\label{eq:FP}
	\left( -\nabla^2+ m^2-s-1 \right) \phi_{\mu_1 \cdots \mu_s} =0 \; , \qquad 	\nabla^\mu \phi_{\mu \mu_2 \cdots \mu_s} = 0\; , \qquad 
	\phi^\lambda_{\;\lambda_{\mu_3} \cdots \lambda_{\mu_s}}	=0 \; . 
\end{align}
Here the physical mass $M$ and dual dimension are given by
\begin{align}
	s(s-3)+M^2 =s(s-3)+m^2 -(s-1)^2 = \Delta (\Delta -2) -s \;. 
\end{align}
The massless case $M=0$ corresponds to $\Delta=s$.  The Fierz-Pauli field has two independent polarizations, which can alternatively be captured in terms of a pair of first-order equations
\begin{align}\label{eq:1storder}
	\epsilon\indices{_{\mu_1}^{\alpha \beta}}\nabla_\alpha \phi_{\beta \mu_2 \cdots \mu_s} =&\mp m \, \phi_{\mu_1 \mu_2 \cdots \mu_s} \; , \qquad m>0 \; . 
\end{align}
One can show that solutions to \eqref{eq:1storder} are solutions to \eqref{eq:FP}. The solutions to the two equations form the complete set of solutions to \eqref{eq:FP} as shown in \cite{Acosta:2021oqt}.

As in section \ref{sec:ScalarSpec} we can convert the Lorentzian wave equation to the Euclidean eigenvalue problem by Wick rotating and sending $m\to i\lambda$
\begin{equation}\label{eq:EuFP}
     -\nabla^2 \phi_{\mu_1 \cdots \mu_s} = \left(\lambda^2 +1 +s  \right)\phi_{\mu_1 \cdots \mu_s}  \; , \qquad 	\nabla^\nu \phi_{\nu \mu_2 \cdots \mu_s} = 0\; , \qquad 
	\phi^\nu_{\;\;\nu {\mu_3} \cdots {\mu_s}}	=0 \; . 
\end{equation}
Similar to the Lorentzian case, this can be formulated in terms of a first order equation
\begin{align}\label{eq:Euc1storder}
	\epsilon\indices{_{\mu_1}^{\alpha \beta}}\nabla_\alpha \phi_{\beta \mu_2 \cdots \mu_s} =&- i\lambda \, \phi_{\mu_1 \mu_2 \cdots \mu_s} \; . 
\end{align}
Note that we do not have the $(\pm)$-branches as in the Lorentzian case \eqref{eq:1storder}, because as we will see, we will take $\lambda \in \mathbb{R}$. The solution to \eqref{eq:Euc1storder} with $\lambda>0$ together with the solution to \eqref{eq:Euc1storder} with $\lambda' = -\lambda<0$ give rise to the two independent solutions to \eqref{eq:EuFP} at a fixed $\lambda^2$.
The Lorentzian problem was solved in \cite{Datta:2011za} in order to derive the higher-spin  QNMs, and we can adapt these results to derive the spectral measure for the massless higher spin fields. The system \eqref{eq:FP} couples the different components of the field, but \cite{Datta:2011za} found linear combinations 
\begin{align}\label{eq:eucansatz}
	\phi_{[p]} = e^{-i \omega_{kl}t +i l\varphi} R_{[p]}(\xi)   \; , \qquad \qquad \omega_{kl}=\frac{2\pi k}{\beta}-il\Omega_H  \; , \qquad k,l\in\mathbb{Z} \; \qquad p=0,1,\cdots, s 
\end{align}
that diagonalize the kinetic term. Using this ansatz the differential equation for the radial functions becomes
\begin{align}
 z(1-z)\partial_z^2R_{[p]} +(1-z)\partial_z R_{[p]}(z) + \left[\frac{k_+^2}{4z}-\frac{k_-^2}{4}+\frac{1-(2p-s+i\lambda)^2}{4(1-z)}\right]R_{[p]}(z)=0 \; , 
 \end{align}
 which is of the form \eqref{eq:scalarSpec} for a shifted $i\lambda\to i\lambda +2p-s$. We can therefore use the results of section \ref{sec:ScalarSpec} to write down the wavefunctions $R_{[p]}$
\begin{equation}
    R_{[p]}(\xi)=N_{[p]}(\tanh\xi)^{|k|}(\text{sech}\xi)^{1+i\lambda+2p-s}\;_2F_1\left(\frac{1+i\lambda+2p-s}{2} -\frac{i}{2}(i|k|+k_-);\frac{1+i\lambda+2p-s}{2}-\frac{i}{2}(i|k|-k_-) ;1+|k|;\tanh^2\xi\right) 
\end{equation}
The relative normalization factor $N_{[p]}$ is fixed by imposing the first order equation \eqref{eq:Euc1storder} to be
\begin{equation}\label{eq:Nfactor}
 N_{[p]}(\lambda)=    (-1)^p \prod_{j=0}^{p-1} \frac{2j-s+1+i\lambda +(|k|-ik_-)}{2j-s+1+i\lambda-(|k|+ ik_-)} \; . 
\end{equation}
Using the connection formula \eqref{eq:2f1 form} we find that near the boundary\footnote{The wavefunctions grow exponentially near the boundary but are still delta function normalizable because the norm involves inverse powers of the metric for spinning fields.}
\begin{equation}\label{eq:TensorFalloff}
    R_{[p]}\sim N_{[p]}(\lambda)C^{\text{in}}_{kl[p]}(\lambda)e^{\xi(s-1-2p)}e^{-i \lambda \xi} + N_{[p]}(\lambda)C^{\text{out}}_{kl[p]}(\lambda) e^{\xi(2p-s-1)}e^{i\lambda \xi}\; , 
\end{equation}
where
\begin{equation}
    C_{kl[p]}^{\text{in}}(\lambda)=C_{kl}(\lambda-i(2p-s)) \; , \qquad \qquad C_{kl[p]}^{\text{out}}(\lambda)=C_{kl}(-\lambda+i(2p-s)) \; , 
\end{equation}
with $C_{kl}$ given by \eqref{eq:cklnew}. In order to define the scattering phase, we have to look for the components that dominate the incoming and outgoing waves. Comparing the exponential growth of the two terms in \eqref{eq:TensorFalloff} one sees that the incoming wave is dominated by the $p=0$ component while the outgoing wave is dominated by the $p=s$ component. Discarding the universal terms which do not contribute to the scattering phase and the Rindler-like contributions, one obtains
\begin{equation}
    \mu_{kl}^{s}(\lambda)
    =\frac{1}{2\pi i}\partial_\lambda \log \frac{N_{[s]}(\lambda)C_{kl}^{ \rm BTZ}(-\lambda+is)}{C_{kl}^{\rm BTZ}(\lambda+is)} \; .  
\end{equation}
Note that this expression is not invariant under $\lambda \to -\lambda$, which is a symmetry of the Euclidean Fierz-Pauli system \eqref{eq:EuFP}. This is because the first-order formulation \eqref{eq:Euc1storder} does not respect this $\mathbb{Z}_2$-symmetry. However, recall that in order for the solutions to \eqref{eq:Euc1storder} to form a complete set of solutions to \eqref{eq:EuFP}, we need to take $\lambda\in\mathbb{R}$. Therefore, while it is not manifest, we do have the $\lambda \to -\lambda$ symmetry in the full space of solutions. When integrating against $\mathbb{Z}_2$-invariant functions of the form $f(\lambda^2)$ over $\lambda\in\mathbb{R}$ (as we do when computing the heat kernel), we can write
\begin{align}
    \int_{-\infty}^\infty d\lambda \, \mu_{kl}^s(\lambda) f(\lambda^2) = \int_{-\infty}^\infty d\lambda \, \frac{\mu_{kl}^s(\lambda)+\mu_{kl}^s(-\lambda)}{2} f(\lambda^2) \; . 
\end{align}
Since
\begin{align}
    \frac{N_{[s]}(\lambda)}{N_{[s]}(-\lambda)} = \frac{C^{\rm BTZ}_{kl}(\lambda+is)}{C^{\rm BTZ}_{kl}(-\lambda+is)} \frac{C^{\rm BTZ}_{kl}(-\lambda-is)}{C^{\rm BTZ}_{kl}(\lambda-is)} B_{kl}(\lambda)\; , \qquad B_{kl}(\lambda) = \frac{\cos \pi \left(|k|+ i k_-\right)+\cos \pi (s+i \lambda)}{\cos \pi \left(|k|+ i k_-\right)+\cos \pi (s-i \lambda)} \; , 
\end{align}
we can then write down the manifestly $\mathbb{Z}_2$-invariant spectral measure
\begin{align}\label{eq:SpinMeasure}
    \frac{\mu_{kl}^s(\lambda)+\mu_{kl}^s(-\lambda)}{2} =\frac{1}{4\pi i}\partial_\lambda \log \left[\frac{C^{\rm BTZ}_{kl}(-\lambda-is)}{C^{\rm BTZ}_{kl}(\lambda+is)} \frac{C^{\rm BTZ}_{kl}(-\lambda+is)}{C^{\rm BTZ}_{kl}(\lambda-is)}B_{kl}(\lambda)\right]=\frac{\mu_{kl}(\lambda+is) + \mu_{kl}(\lambda-is)}{2} \; , 
\end{align}
where we have used $\partial_\lambda \log B_{kl}=0$ and  $\mu_{kl}(\lambda)$ is the scalar spectral density \eqref{eq:mukl}.  Now we turn to the heat kernel
\begin{align}\label{eq:HSheatkerneldef}
     K(T) \equiv \text{Tr} \, e^{-\left(-\nabla_s^2 \right) T} =  \int_\mathbb{R} d\lambda \, \frac{\mu(\lambda+is) + \mu(\lambda-is)}{2}  \,  e^{-(\lambda^2+s+1)T}\; . 
\end{align}
 Naively, one might compute the quantity
\begin{equation}
    \int d\lambda \, \mu (\lambda\pm is) \, e^{-(\lambda^2+s+1)T} 
\end{equation}
using the scalar spectral measure \eqref{eq:ScalarMeasure}, together with the integral
\begin{equation}
    \int_{-\infty}^\infty \cos(2\pi p [\lambda\pm is] \tau_2) e^{-(\lambda^2+s+1)T}d\lambda=\sqrt{\frac{\pi}{T}}e^{-(s+1)T-(\pi p \tau_2)^2/T}\cosh \left(2\pi p s \tau_2\right) \; . 
\end{equation}
This results in  a ``bulk heat kernel''
\begin{align}\label{eq:bulkheatkernel}
    K(T)_{\text{bulk}}=\tau_2\sum_{p=1}^\infty\frac{\cosh{\pi p s \left(\frac{2\pi}{\beta_L}+\frac{2\pi}{\beta_R}\right)}}{\sinh{\pi p \frac{2\pi}{\beta_L}}\sinh{\pi p \frac{2\pi}{\beta_R}}}\sqrt{\frac{\pi}{T}}e^{-(s+1)T-(\pi p \tau_2)^2/T} \; , 
\end{align}
which differs from the heat kernel obtained in \cite{David:2009xg}. Nonetheless, let us use this naive heat kernel to compute the massive spin-$s$ determinant \eqref{eq:ZPImass}, i.e. 
\begin{align}\label{eq:BTZPIspin}
	\log Z^{s,\Delta}_{\rm PI} \Big|_{\rm naive}=& \int_0^\infty \frac{dT}{2T}  e^{-\left( \Delta (\Delta -2)-s\right) T}	K(T)_{\rm bulk}\; . 
\end{align}
Computing the $T$-integral using
\begin{equation}
    \sqrt{\pi}\int_0^\infty\frac{dT}{T}\frac{e^{-(\Delta-1)^2T-(\pi p \tau_2)^2/T}}{\sqrt{T}}=\frac{e^{-2\pi p \tau_2(\Delta-1)}}{p\tau_2} 
\end{equation}
one would find 

\begin{align}\label{eq:logZbulk}
 \log Z^{s,\Delta}_{\rm PI} \Big|_{\rm naive} = \frac12\sum_{p=1}^\infty\frac{1}{p} \frac{e^{-2\pi p \tau_2(\Delta-1)}}{|\sin p\pi \tau|^2}\cosh \left(2\pi p s \tau_2\right)=\sum_{p=1}^\infty\frac{1}{p} \frac{(q_Lq_R)^{p(\Delta-1)/2}}{(q_L^{p/2}-q_L^{-p/2})(q_R^{p/2}-q_R^{-p/2})}\left((q_Lq_R)^{ps/2}+(q_Lq_R)^{-ps/2}\right) 
\end{align}
which is the bulk partition function \eqref{eq:masspinbulk} derived from the naive DHS formula without the edge subtractions; in other words, we have $Z^{s,\Delta}_{\rm PI} \Big|_{\rm naive}  = Z_{\rm bulk}$. This justifies calling \eqref{eq:bulkheatkernel} a ``bulk heat kernel."

To understand what went wrong, we have to treat more carefully the spectral density. The representation of \eqref{eq:SpinMeasure} 
\begin{equation}\label{eq:SpinningMeasure}
    \mu^s(\lambda)= \frac{1}{4\pi i }\sum_{k,l\in \mathbb{Z}}\sum_{\pm}\sum_{n=0}^\infty \left(\frac{1}{\lambda+is+\lambda_{nkl}^\pm}-\frac{1}{\lambda+is-\lambda_{nlk}^\pm}+\frac{1}{\lambda-is+\lambda_{nkl}^\pm}-\frac{1}{\lambda-is-\lambda_{nlk}^\pm}\right) 
\end{equation}
continues to hold, but the representation \eqref{eq:ScalarMeasure} has to be altered. In particular, when $\Im a<0$, we have
\begin{align}\label{eq:normalInt}
    \frac{1}{i}\left(\frac{1}{\lambda + a} - \frac{1}{\lambda -a}\right) = \int_0^\infty dt \left(e^{-i(\lambda+ a) t}+e^{i(\lambda- a) t} \right)  = \int_0^\infty dt \left(e^{i\lambda  t}+e^{-i\lambda  t} \right)e^{-i a t} \; . 
\end{align}
When $\Im a>0$, we have instead
\begin{align}\label{eq:flipInt1}
    \frac{1}{i}\left(\frac{1}{\lambda + a} - \frac{1}{\lambda -a}\right) 
    =& -\frac{1}{i}\left(\frac{1}{-\lambda - a} - \frac{1}{-\lambda +a}\right) \nn\\
    =&-\int_0^\infty dt \left(e^{-i(-\lambda - a) t}+e^{i(-\lambda +a) t} \right) \nn\\
    =&-\int_0^\infty dt \left(e^{i\lambda  t}+e^{-i\lambda  t} \right)e^{i a t} \; . 
\end{align}
\eqref{eq:flipInt1} differs from \eqref{eq:normalInt} by a sign flip $a\to -a$ together with an overall minus sign. More explicitly, when $\Im a>0$,
\begin{align}\label{eq:flipInt}
    \frac{1}{i}\left(\frac{1}{\lambda + a} - \frac{1}{\lambda -a}\right) 
    = \int_0^\infty dt \left(e^{i\lambda  t}+e^{-i\lambda  t} \right) \left[\underbrace{e^{-i a t}}_{\rm naive}+\underbrace{\left(-e^{-i a t}-e^{i a t} \right)}_{\rm correction}\right]\; . 
\end{align}
In order to find the correct version of the representation \eqref{eq:ScalarMeasure}, we need to compute the correction term in  \eqref{eq:flipInt} for each pole whose imaginary part changes sign, and then add it to the naive formula \eqref{eq:ScalarMeasure}.

The terms which need to be corrected are the first and last term in \eqref{eq:SpinningMeasure} that have
\begin{align}\label{eq:imabound}
		\Im \left( \lambda^\pm_{nkl} +i s \right) =s - \left(1+2n+|k| \right) \geq 0 \; , 
\end{align}
or equivalently
\begin{align}\label{eq:ineq}
n\leq \frac{s-1-|k|}{2} \; . 
\end{align}
In deriving this formula it is important to recall that $T_L$ and $T_R$ are complex conjugates, so that
\begin{equation}
    \text{Re}\left(\frac{T}{T_{L/R}}\right)=\frac{1}{2} \left(\frac{T}{T_{ L}}+\frac{T}{T_{ R}}\right)=T\cdot\frac{\beta_{ L}+\beta_{ R}}{2}=1\;. 
\end{equation}
Similarly the real parts of the $\lambda_{nkl}^\pm$ are given by 
\begin{equation}\label{eq:lambdareal}
    \Re{\lambda_{nkl}^\pm}=\pm k(i\Omega) \mp \frac{l}{r_+} \; .
\end{equation}
When \eqref{eq:ineq} is a strict inequality $<$, we can write down the correction terms using \eqref{eq:flipInt}. When the equality holds, i.e. $ \lambda^\pm_{nkl}+is$ is real, we would have  $ \lambda^+_{nkl}+is=- (\lambda^-_{nkl}+is)$, and therefore the would-be poles cancel out 
\begin{align}
    & \left(\frac{1}{\lambda+is+ \lambda^+_{nkl}}-\frac{1}{\lambda-is- \lambda^+_{nkl}} \right)+ \left( \frac{1}{\lambda+is+ \lambda^-_{nkl}}-\frac{1}{\lambda-is- \lambda^-_{nkl}}\right) \nn\\
    =& \left(\frac{1}{\lambda+is+ \lambda^+_{nkl}}-\frac{1}{\lambda-is- \lambda^-_{nkl}}\right)+ \left( \frac{1}{\lambda+is+ \lambda^-_{nkl}}-\frac{1}{\lambda-is- \lambda^+_{nkl}} \right)\nn\\
    =& \, 0  \; . 
\end{align}
In this case the second correction term in \eqref{eq:flipInt} is not needed, so 
\begin{equation}\label{eq:muedgedef}
    \mu_{\text{edge}}(\lambda)=\frac12\int_0^\infty \frac{dt}{2\pi}\left(e^{-i\lambda t}+e^{i\lambda  t} \right)\sum_{l\in \mathbb{Z}}\sum_{\pm} \sum_{k=-(s-1)}^{s-1}\sum_{n=0}^{\lfloor\frac{s-1-|k|}{2}\rfloor}\left(e^{-i[\lambda_{nkl}^\pm+is]  t}+e^{i[\lambda_{nkl}^\pm+is]  t} (1-\delta_{n,\frac{s-1-|k|}{2}})\right)\; . 
\end{equation}
Since $\Re{\lambda^+_{nkl}}=-\Re{\lambda^-_{nkl}}$ and $\Im{\lambda^+_{nkl}}=\Im{\lambda^-_{nkl}}$ we have
\begin{align}
    &\sum_{l\in \mathbb{Z}}\sum_{\pm} \sum_{k=-(s-1)}^{s-1}\sum_{n=0}^{\lfloor\frac{s-1-|k|}{2}\rfloor}\left(e^{-i[\lambda_{nkl}^\pm+is]  t}+e^{i[\lambda_{nkl}^\pm+is]  t} (1-\delta_{n,\frac{s-1-|k|}{2}})\right)\nn\\
    =& \sum_{l\in \mathbb{Z}}\sum_{\pm} \sum_{k=-(s-1)}^{s-1}\sum_{n=0}^{\lfloor\frac{s-1-|k|}{2}\rfloor}\left(e^{-i[\lambda_{nkl}^\pm+is]  t}+e^{i[\lambda_{nkl}^\mp+is]  t} (1-\delta_{n,\frac{s-1-|k|}{2}})\right)\nn\\
    =& \sum_{l\in \mathbb{Z}}\sum_{\pm} \sum_{k=-(s-1)}^{s-1}\sum_{n=0}^{\lfloor\frac{s-1-|k|}{2}\rfloor}e^{-i \Re{\lambda^\pm_{nkl}}  t }\left(e^{\Im[\lambda_{nkl}^\pm+is]  t}+e^{-\Im[\lambda_{nkl}^\pm+is]  t} (1-\delta_{n,\frac{s-1-|k|}{2}})\right)\;. 
\end{align}
Noting also that $\Re{\lambda^\pm_{nkl}}$ is independent of $n$, we can use the identity
\begin{align}\label{eq:flipsumformula}
\sum_{n=0}^{L-1}  q^{L-2n-1}
    =\sum_{n=0}^{\lfloor \frac{L-1}{2}\rfloor} \left( q^{L-2n-1}+q^{1+2n-L} (1-\delta_{n,\frac{L-1}{2}} )\right)\;  
\end{align}
to write
\begin{equation}
    \mu_{\text{edge}}(\lambda)=\frac12\int_0^\infty \frac{dt}{2\pi}\left(e^{-i\lambda t}+e^{i\lambda t} \right)e^{st} \sum_{k=-(s-1)}^{s-1}\sum_\pm\sum_{n=0}^{s-1-|k|}\sum_{l\in \mathbb{Z}}e^{-i\lambda_{nkl}^\pm t}\; . 
\end{equation}
Now
\begin{align}
\sum_{k=-(s-1)}^{s-1}&\sum_\pm\sum_{n=0}^{s-1-|k|}\sum_{l\in \mathbb{Z}}e^{-i\lambda_{nkl}^\pm t}
=2\pi r_+\sum_{p\in \mathbb{Z}} \delta(t-2\pi r_+ p)\sum_{k=-(s-1)}^{s-1}\left[\sum_{n=0}^{s-1-|k|}e^{-(2n+1)t}\right]
\left(
e^{-(|k|-k\Omega_H)t}+e^{-(|k|+k\Omega_H)t}
\right)\nn\\
    &=2\pi r_+\sum_{p\in \mathbb{Z}} \delta(t-2\pi r_+ p)\sum_{k=-(s-1)}^{s-1}\left(
e^{-|k|(1-\Omega_H)t}+e^{-|k|(1+\Omega_H)t}
\right)\sum_{n=0}^{s-1-|k|}e^{-(2n+1)t} \; , 
\end{align}
where in the second line we swapped the terms with $k<0$.  Performing the sums and doing the $t$ integral we get
\begin{equation}
    \mu_{\text{edge}}(\lambda)=2r_+\sum_{p=1}^\infty\cos(2\pi p r_+ \lambda)\frac{(1-q_L^{-ps})(1-q_R^{-ps})}{(1-q_L^{-p})(1-q_R^{-p})}(q_Lq_R)^{p\frac{s-1}{2}} \; . 
\end{equation}
 Let us now include the correction to the heat kernel \eqref{eq:HSheatkerneldef}.
Again making use of the integral \eqref{eq:Tintegralformula},
\begin{equation}
    \int_{-\infty}^\infty \cos(2\pi p \lambda \tau_2) e^{-(\lambda^2+s+1)T}d\lambda=\sqrt{\frac{\pi}{T}}e^{-(s+1)T-(\pi p \tau_2)^2/T} \;  
\end{equation}
we find an ``edge heat kernel"
\begin{align}\label{eq:edgeheatkernel}
    K(T)_{\text{edge}}&=2\tau_2\sum_{p=1}^\infty\frac{(1-q_L^{-ps})(1-q_R^{-ps})}{(1-q_L^{-p})(1-q_R^{-p})}(q_Lq_R)^{p\frac{s-1}{2}}\sqrt{\frac{\pi}{T}}e^{-(s+1)T-(\pi p \tau_2)^2/T}\;   \nn\\
     & =2\tau_2\sum_{p=1}^\infty\frac{\sinh{\pi p s \frac{2\pi}{\beta_L}}}{\sinh{\pi p \frac{2\pi}{\beta_L}}}\frac{\sinh{\pi p s \frac{2\pi}{\beta_R}}}{\sinh{\pi p \frac{2\pi}{\beta_R}}}\sqrt{\frac{\pi}{T}}e^{-(s+1)T-(\pi p \tau_2)^2/T} \; . 
\end{align}
 It is then straightforward to compute the full heat kernel,
\begin{align}\label{eq:fullheatkernel}
    K(T)= K(T)_{\text{bulk}} -K(T)_{\text{edge}} = \tau_2\sum_{p=1}^\infty\frac{\cosh{\pi p s \left(\frac{2\pi}{\beta_L}-\frac{2\pi}{\beta_R}\right)}}{\sinh{\pi p \frac{2\pi}{\beta_L}}\sinh{\pi p \frac{2\pi}{\beta_R}}}\sqrt{\frac{\pi}{T}}e^{-(s+1)T-(\pi p \tau_2)^2/T} \; , 
\end{align}
 agreeing precisely with the result obtained in \cite{David:2009xg} by the method of images.  Integrating \eqref{eq:fullheatkernel} produces the full massive spin-$s$ partition function \eqref{eq:ZPImassresult}. One can also directly compute the edge partition function using the edge heat kernel 
 \begin{align}
	\log Z_{\text{edge}} =& \int_0^\infty \frac{dT}{2T}  e^{-\left( \Delta (\Delta -2)-s\right) T}	K(T)_{\rm edge}\; . 
\end{align}
Using
\begin{equation}
    \sqrt{\pi}\int_0^\infty\frac{dT}{2T}\frac{e^{-(\Delta-1)^2T-(\pi p \tau_2)^2/T}}{\sqrt{T}}=\frac{e^{-2\pi p \tau_2(\Delta-1)}}{2p\tau_2} 
\end{equation}
we have
\begin{equation}\label{eq:logZedge}
    \log Z_{\text{edge}}=\sum_{p=1}\frac{1}{p}\frac{(1-q_L^{-ps})(1-q_R^{-ps})}{(1-q_L^{-p})(1-q_R^{-p})}(q_Lq_R)^{p(s-1)/2}(q_Lq_R)^{p(\Delta-1)/2} \; , 
\end{equation}
which is \eqref{eq:masspinedge}. Subtracting \eqref{eq:logZedge} from \eqref{eq:logZbulk} yields the correct massless higher spin determinant when $\Delta=s$.

\section{Conclusion and implications for Kerr}\label{sec:Conclusion}

The low temperature scaling of the black hole partition function $Z_{BH}\sim T^{3/2}e^{S_0}$ 
resolves several old puzzles about cold black holes, but the source of the effect is quite subtle.  The existing derivations, which only make use of fluctuations in the near-horizon region, obtain the effect from path integrals over pure diffeomorphisms with noncompact support in the infinite throat. Black holes at finite temperature have finite throats, and the completion of these throat diffeomorphisms in the full black hole geometry is not currently known. 

In this work we took a step towards proving the $T^{3/2}$ scaling in a way that does not involve a near-horizon calculation.  Using the DHS representation of the BTZ one-loop path integral, we identified the particular branch of QNMs that are responsible for the $T^{3/2}$ behavior and isolated the relevant characteristics of the spectrum which might produce the same result in Kerr. 
As a byproduct of our analysis, we obtained new formulas for the spectral measure for fields of any mass and integer spin in Euclidean BTZ. These explicit formulas allowed us to prove the DHS formula (without assumptions) and to rederive the BTZ heat kernel using an integral over eigenvalues rather than the method of images. We believe that this approach will be useful for studying the effect in more realistic black holes, since it makes clear when the DHS assumptions might fail and provides the relevant corrections. 

The spectral density calculations highlight another reason that the DHS approach seems likely to make contact with the near-horizon analyses. Although the factor of $T^{3/2}$ arises from discrete zero modes in AdS$_2$/NHEK, most of the spectrum in these geometries is actually continuous (and contributes known logarithmic corrections to the extremal entropy \cite{Sen:2014aja}). Indeed, it is somewhat atypical to have so many discrete eigenvalues on a noncompact infinite-volume geometry. An exception obviously occurs for spinning fields in AdS$_2$, as noted in \cite{Camporesi:1994ga}. However, these authors also showed that there are no graviton discrete modes in hyperbolic space $\mathbb{H}_n$ for $n>2$, and the authors of \cite{Acosta:2021oqt} make a similar claim for gravitons on Euclidean BTZ. The fact that we are able to reproduce the heat kernel (and $T^{3/2}$) using an integral over eigenvalues with no contribution from a point spectrum seems to support this claim. It would be very interesting to investigate the analogous questions for higher dimensional black holes. In other words, does $T^{3/2}$ always come from the essential spectrum when the full black hole geometry is used?

At least for the BTZ black hole, the mechanism producing $T^{3/2}$ in the full geometry seems to be quite different from the mechanism in the throat calculation.  We effectively replace the discrete product of eigenvalues in the throat by a discrete product of  QNMs in BTZ. In other words, \textit{resonances replace eigenfunctions}, which is often the case for many questions on infinite-volume manifolds \cite{BorthwickBook2016,dyatlov2019mathematical}. 
 Even when the spectrum of $-\nabla^2$ on eigenfunctions is continuous, the spectrum of resonances is typically discrete.
The next obvious step is to repeat the analysis for Kerr and Reissner-Nordstrom, attempting to leverage the analytic results on the QNM spectrum in the near-extremal limit \cite{Detweiler:1980gk,Sasaki:1989ca,Glampedakis:2001fg,Ferrari:1984zz,Cardoso:2004hh,Hod:2008zz,Hod:2012bw,Yang:2012he,Yang:2012pj,Yang:2013uba,Cook:2014cta,Dias:2015wqa}.

\subsection*{Acknowledgements}
We thank Cindy Keeler for discussions. DK is supported by  DOE grant de-sc/0007870. AL is supported by the Stanford Science Fellowship. The work of CT is supported by the Marie Sklodowska-Curie Global Fellowship (ERC Horizon 2020 Program) SPINBHMICRO-101024314. CT also acknowledges support from the Simons Center for Geometry and Physics, Stony Brook University at which some of the research for this paper was performed.

\appendix

\section{Log corrections from massless higher spin fields }\label{app:HS}

The reduced-determinant analog of \eqref{eq:gravPI} for massless spin-$s$ fields was derived in \cite{Gaberdiel:2010ar} and takes the form
\begin{align}\label{eq:HSPI}
	Z_{\rm PI} = \left( \frac{ \det \left( -\nabla_{s-1}^2 +s^2-s \right) }{ \det \left( -\nabla_{s}^2 - 3s+s^2\right)  } \right)^{1/2} \;. 
\end{align}
As usual, this formula assumes a specific analytic continuation of the contour in field space that corrects the sign of the kinetic terms for certain components of the higher spin field. \eqref{eq:HSPI} can be computed by considering limits of the massive spin-$s$ results in section \ref{sec:massHS}:
\begin{align}\label{eq:HSlimits}
	s -1 : \qquad \Delta \to s+1 \;, \qquad s: \qquad \Delta \to s   \; . 
\end{align}
According to formula \eqref{eq:ZPImassresult}, after removing the edge-mode contribution the two determinants take the form
\begin{align}
	-\frac12 \log  \det \left( -\nabla_{s}^2 - 3s+s^2\right)   &= -\sum_{n_L ,n_R =0}^\infty \log \left( 1- q_L^{s+n_L}q_R^{n_R}\right) \left( 1- q_L^{n_L}q_R^{s+n_R}\right) \;,\\
 	-\frac12 \log \det \left( -\nabla_{s-1}^2 +s^2-s \right) &= -\sum_{n_L ,n_R =0}^\infty \log \left( 1- q_L^{s+n_L}q_R^{1+n_R}\right) \left( 1- q_L^{1+n_L}q_R^{s+n_R}\right) \;. 
\end{align}
Combining the two terms, we obtain
\begin{align}
	\log Z_{\rm PI} = \sum_{p=1}^\infty \frac1p \frac{q_L^{ps}(1 -q_R^p) + q_R^{ps} (1-q_L^p) }{(1-q_L^p)(1-q_R^p)}  \;. 
\end{align}
Expanding out and simplifying one finds
\begin{eqnarray}
	\log Z_{\rm PI} & = & \sum_{n_L,n_R=0}^\infty \sum_{p=1}^\infty \frac1p ( q_L^{ps}(1 -q_R^p) + q_R^{ps} (1-q_L^p) ) q_L^{p n_L} q_R^{p n_R}   \nonumber \\
	 &= & \sum_{n_L,n_R=0}^\infty \sum_{p=1}^\infty \frac1p ( q_L^{p(s+n_L)}q_R^{p n_R}  - q_L^{p(s+n_L)} q_R^{p(n_R+1)} + q_R^{p(s+n_R)} q_L^{p n_L} - q_R^{p(s+n_R)} q_L^{p(n_L+1)} )   \nonumber \\
   & = &  \sum_{p=1}^\infty \frac1p \left( \sum_{n_L=s,n_R=0}^{\infty} q_L^{p n_L}q_R^{p n_R}  - \sum_{n_L=s,n_R=1}^{\infty} q_L^{p n_L} q_R^{p n_R} + \sum_{n_L=0,n_R=s}^{\infty} q_R^{p n_R} q_L^{p n_L} - \sum_{n_L=1,n_R=s}^\infty q_R^{p n_R} q_L^{p n_L} \right) \nonumber \\
   & = &  \sum_{p=1}^\infty \frac1p \left( \sum_{n_L=s}^{\infty} q_L^{p n_L} + \sum_{n_R=s}^{\infty} q_R^{p n_R}  \right) \; . 
\end{eqnarray}
Using $- \log (1-x) = \sum_{p=1}^{\infty} \frac{x^p}{p}$ we get 
\begin{equation}
   \log Z_{\rm PI} = - \sum_{n_L =s}^{\infty}  \log (1-q_L^{n_L}) - \sum_{n_R =s}^{\infty}  \log (1-q_R^{n_R}) 
\end{equation}
and hence
\begin{equation}
   Z_{\rm PI} = \prod_{n_L,n_R =s}^{\infty}  \frac{1}{(1-q_L^{n_L})} \frac{1} {(1-q_R^{n_R})}\;. 
\end{equation}
Taking the low-$T$ limit as in \eqref{eq:BTZprod} we have, using zeta function regularization, 
\begin{equation}\label{eq:HSlogT}
   Z_{\rm PI} \, \propto  \, T^{s-\frac12} \; . 
\end{equation}
Models with families of higher spin fields will receive multiple such contributions. For instance, for higher-spin gravity with $SL(n)\times SL(n)$ as the higher-spin group, which has massless fields of spin $s=2,3\cdots, n,$ the coefficient of log $T$ will be $(n^2-1)/2$, which is the dimension of the higher spin group divided by 4 \cite{Datta:2021efl}. Models with higher spins producing factors like \eqref{eq:HSlogT} have also recently been considered in nearly-AdS$_2$ throat calculations \cite{Gonzalez:2018enk,Kruthoff:2022voq}.

\bibliographystyle{apsrev4-1long}
\bibliography{Bib.bib}
\end{document}